\documentclass{ISMA_USD}
\usepackage[sorting=none]{biblatex}
\usepackage{xcolor}
\DeclareUnicodeCharacter{0302}{*************************************}

\hypersetup{pdftitle  = {Bayesian NVH metamodels to assess interior cabin noise using measurement databases},
	pdfauthor = {V. Prakash, O. Sauvage, J. Antoni, L. Gagliardini},
	pdfkeywords = {Automotive NVH, Interior cabin noise, Bayesian metamodel, Uncertainty quantification, MCMC, PyMC3}}

\title{Bayesian NVH metamodels to assess interior cabin noise using measurement databases}

\author[1,2]{V. Prakash}
\author[1]{O. Sauvage}
\author[2]{J. Antoni}
\author[3]{L. Gagliardini}
  
\affil[1]	{Stellantis N.V., Automotive Research \& Advanced Engineering department, 
	\NewLineAffil OpenLab Vibro-Acoustic-Tribology@Lyon, France  \NewLineAffil 
	e-mail: \textbf{vinay.prakash@stellantis.com}\NewAffil}

\affil[2]	{Univ Lyon, INSA Lyon, LVA, EA677, 69621 Villeurbanne, France \NewAffil}
\affil[3]   {Stellantis N.V., NVH Department, Velizy-Villacoublay, France}
		
\date{}

\begin{document}



\abstract{In recent years, a great emphasis has been put on engineering the acoustic signature of vehicles that represents the overall comfort level for passengers. Due to the inherent uncertain behavior of production cars, probabilistic metamodels or surrogates can be useful to estimate the NVH dispersion and assess different NVH risks. These metamodels follow physical behaviors and shall aid as a design space exploration tool during the early stage design process to support the NVH optimization. The measurement databases constitute different noise contributions such as aerodynamic noise (wind-tunnel test), tire-pavement interaction noise (rolling noise), and noise due to electric motors (whining noise).	
This research work proposes a global NVH metamodeling technique for broadband noises such as aerodynamic and rolling noises exploiting the Bayesian framework that takes into account the prior (domain-expert) knowledge about complex physical mechanisms. Generalized additive models (GAMs) with polynomials and Gaussian basis functions are used to model the dependency of sound pressure level (SPL) on predictor variables. Moreover, parametric bootstrap algorithm based on data-generating mechanism using the point estimates is used to estimate the dispersion in unknown parameters. Probabilistic modelling is carried out using an open-source library (PyMC3) that utilizes No-U-Turn sampler (NUTS) and the developed models are validated using cross-validation technique.}

\maketitle


\section{Introduction}

Noise, vibration and harshness (NVH) characteristics of a vehicle is an important criterion for validation during the development phase as it significantly affects the customer's quality perception and the overall image of the vehicle. For automotive OEMs, a significant challenge is to continuously develop new methods for estimating the performance indicators and carry out large number of testing campaigns which pushes them towards leveraging simulation-driven design processes. During the vehicle development phase, the vibration response and acoustic levels can be determined using physics-based numerical simulations considering the complex full-vehicle structural-acoustic computational models (refined 3D finite element models), which are usually time consuming. The highly uncertain behavior that arises from manufacturing tolerances, natural variability in material properties and conditions employed during the physical testing procedures \cite{durand2008structural}, leads to a challenging level of uncertainty from a modeling perspective. In such cases, methods based on time-consuming physics based simulations are not relevant as no precise (unique) design nor the detailed information about the vehicle and the powertrain is available. As a result, a comprehensive framework for vehicle NVH assessment with domain expertise and experimental databases is needed along with fast computing models. Hence, the early-stage design aspects raise the following questions: First, how much information should be considered available to the designer in order to derive useful conclusions? Second, how can the available measurement databases be exploited quantitatively to provide the best relevant prior knowledge together with a measure of uncertainty in the outputs?

In light of these challenges, so called ``metamodels" or ``surrogate" models can be used to replace the computationally intensive simulation models or measurement data (retrieved through physical testing), using analytical relations between the design variables and system responses. Metamodels make it possible to quickly explore the design alternatives through parameterized calculations when the general relative performance of design alternatives is more important than a precise estimate. Conceptualized in 1974 by Blanning and popularized by Kleijnen as a ``model of a model" \cite{kleijnen1975comment}, metamodels have been used as ``cheap" yet robust approximations to get better insight into the functional relationships. Metamodelling techniques have been applied to many engineering disciplines such as crashworthiness, engine modelling, structural reliability, NVH and so on \cite{kianifar2020performance}. In the context of automotive NVH, metamodels based on design of experiments were employed to achieve minimal piston noise \cite{hoffman_robust_2003}, response surface method was used to minimize the sound pressure level inside the cabin \cite{zheng_design_2015}, radial basis functions was used to reduce the vehicle mass with vibration constraints \cite{kiani_comparative_2016}, and Park et al \cite{park_efficient_2020} reduced the structure-borne noise using Kriging surrogate models. Kuznar et al. \cite{kuvznar2012improving} proposed a novel approach to learning regression models that are consistent with domain-expert knowledge in the form of quantitative constraints for aerodynamic wind-noise. In several other studies, researchers compared and benchmarked different techniques as shown in \cite{chen_multi-objective_2018, ibrahim_surrogate-based_2020}. A detailed review of different metamodelling techniques can be found in \cite{wang2007review}.

In order to quantify uncertainties, deterministic metamodels are not sufficient and instead probabilistic metamodels are used where the output response is not just a point estimate but a probability distribution. This provides flexibility to evaluate design alternatives with a given level of knowledge and metamodel complexity. Monte Carlo (MC) simulation techniques have been primarily used by researchers for the probabilistic quantification of uncertainties. In the field of computational mechanics, several contributions have been made by researchers, which are detailed in \cite{soize2017uncertainty}. More precisely, in the automotive industry, Durand et al. built a nonparametric model to capture the variability in the booming noise prediction through random matrices \cite{durand2008structural}, Barillon et al. proposed a methodology to quantify the variability in booming noise and body-in-white \cite{barillon2012vibro}, and recently, Brogna et al. \cite{brogna2018probabilistic} used Bayesian approaches with Gibbs sampling to model the global vibro-acoustic behaviour. In this regard, Bayesian approach towards metamodelling suits well as it allows to include prior knowledge (based on domain expertise) about the parameters of the system under consideration.

In this work, a global metamodelling technique is presented, which is based primarily on the parameters that are selected at early stages of the design process. \secref{sec_stochastic_metamodelling} describes the proposed Bayesian metamodelling workflow with the relevant theoretical foundations. The application of developed methods to assess interior cabin noise in vehicles is described in \secref{sec_application_context}. \secref{sec_model_convergence} describes the model assessment with convergence statistics and the conclusions are presented in \secref{sec_conclusions}.

\section{Stochastic metamodelling for uncertainty quantification} \label{sec_stochastic_metamodelling}

\subsection{Proposed workflow}
\begin{figure}[!h]
	\centering
	\includegraphics[scale=0.4]{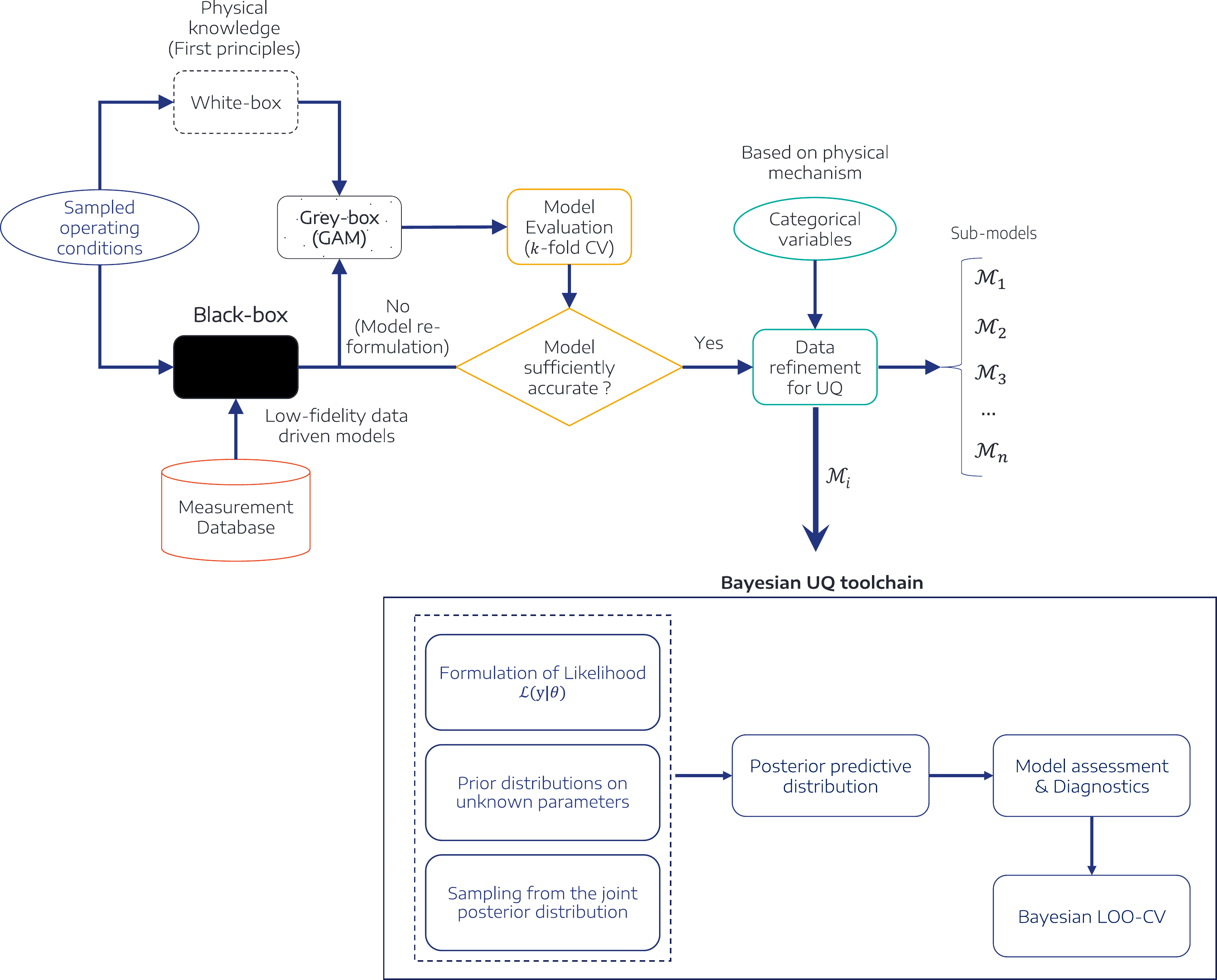} 
	\caption{Flowchart of the proposed metamodelling framework}
	\label{fig_workflow}
\end{figure}

Here, a general workflow for stochastic metamodelling is proposed taking into account the predominant physical laws. \figref{fig_workflow} describes the different steps involved in the workflow. Operating point (OP) conditions are typically the client usage profiles or driving conditions that are collected and represented in the form of a distribution function. These OP conditions usually consist of vehicle speed and the wheel torque which marks the starting point of the metamodelling toolchain. Then, the laws governing the physical process is identified using ``first-principles" (white-box) and are combined with the low-fidelity data-driven regression fit representing the black-box part of the model. This forms an additive functional (grey-box) modelled using Generalized additive models (GAMs) \cite{hastie2017generalized} of two (or multiple) predictor variables. The model is evaluated using $K$-fold cross-validation. To quantify the uncertainties, selection parameters are used to refine the data pertaining to a specific vehicle type. This allows us to consider different dataset for the same model (for instance, the data used for $\mathcal{M}_1$ will be different from the data used for $\mathcal{M}_2$) on the basis of different selection parameters and the Bayesian learning method is applied on one such sub-dataset. Bayesian part of metamodelling is described in \secref{sec_Bayesian}.

\subsection{Generalized Additive Models } \label{sec_GAM}
In this research work, Generalized Additive Models (GAMs) \cite{hastie2017generalized} are exploited for modelling the deterministic surrogates for aerodynamic noise with respect to multiple parameters. In this paper, a lower-case character denotes a scalar variable, a lower-case bold character represents a vector, and a matrix is expressed as an upper-case bold character.

Generalized Linear Model (GLM) relaxes the assumptions of a classical linear model by considering the monotone transformation of the mean response to be a linear function of the predictors and that the conditional distribution of the response belongs to a one-parameter exponential family of densities rather than just being Gaussian \cite{ruppert2003semiparametric}. GAM is an extension to GLM with possibly nonlinear predictors where the smooth functions of predictors are considered in an additive fashion. Let us assume that the observations $(y_i, x_{1i},x_{2i},..,x_{{n_p}i}),~ i=1,2,..,N$ are given, where $y_i$ is the response variable from the vector of observed responses, $\boldsymbol{y} \in \mathbb{R}^{N}$, $N$ being the total number of observed samples and $(x_1,x_2, .., x_{n_p})$ represent different predictor variables from their respective vector quantities $(\boldsymbol{x}_1,\boldsymbol{x}_2,..,\boldsymbol{x}_{n_p}) \in \mathbb{R}^{n_p \times N}$ with $n_p$ being the total number of predictor variables. The general form of the one-parameter exponential family of densities is given by $p(y_i|x_{1i},x_{2i},..,x_{{n_p}i}) = \text{exp}\big(\frac{y_i \theta_i - b(\theta_i)}{\gamma} + c(y_i,\gamma_i)\big)$, where $b(\cdot), c(\cdot), \theta_i$ and $\gamma$ determine the respective distributions. The exponential family includes many distributions for practical modelling, such as the Gaussian, Poisson, binomial, and gamma. 

The additive problem can be formulated for a particular observed point as
\begin{equation}
	y_i = \mu_i + \eta_i, \label{eqn_additive_problem}
\end{equation}
where $\eta_i$ represents the fitting error and the model error, which is supposed to follow $\mathcal{N}(0,\sigma^2)$. Considering $\alpha$ to be the intercept (or bias) term, the conditional expectation of $y_i$, denoted as $\mu_i$ depends on the predictor variables and is given by,
\begin{align}
	\mathbb{E}[y_i] &= \mu_i = \alpha + f_1 (x_{1i}) + f_2(x_{2i}) + f_3(x_{3i}, x_{4i})+...
\end{align}
where $f_j$ denotes a series of smooth functions of the predictor variables $(x_{1i}, x_{2i}, ..,x_{n_p})$. Each of these functions can be approximated with finite basis expansions such as:
\begin{equation}
	f_1(x_1) = \sum_{j=1}^{m}\beta_{1j} {\phi_{1j}(x_1)};~ f_2(x_2) = \sum_{k=1}^{n}\beta_{2k} {\phi_{2k}(x_2)}
\end{equation}
where $\phi(x)$ are the basis expansion functions, $\beta$ the unknown parameter values, $m$ and $n$ the total number of bases in each function.

Re-writing \eqref{eqn_additive_problem} using vector notations leads to:
\begin{equation}
	\boldsymbol{y} = \bX\btheta+\boldsymbol{\eta} \label{eq:state_space_deterministic}
\end{equation}
where $\bX = (\boldsymbol{1, x_1, x_2})$ and $\btheta^{T} = (\alpha, \boldsymbol{\beta_{1}}, \boldsymbol{\beta_{2}})$

Most commonly, these unknown functions $f_1$ and $f_2$ are fitted using penalized splines \cite{ruppert2003semiparametric}. For this study, the focus is on considering the \textit{white}-box knowledge available in the form of analytical equations. The function $f_1$ models these known analytical relations. Therefore, two different deterministic formulations for $f_2$ are considered depending on the choice of basis functions. Also, it is important to build models where the parameters have physical sense and are interpretable which in turn allows more control and flexibility over the output response. Therefore, parametric surrogate modelling approach is investigated, where we have a fixed number of parameters that do not grow with the size of the input data set. Such parametric models make stronger assumptions on the nature of data distribution and are generally faster than the non-parametric models which are more flexible but often computationally intractable \cite{murphy_machine_2012}.

\begin{itemize}
	\item Deterministic model 1: with polynomial basis functions\\
	The first model proposed is with polynomial bases due to their simple structure that comes at a low computational cost where over-fitting issues can also be dealt with depending upon the particular order \cite{myers1989response}. This can be represented as
	\begin{equation}
		f_2(x_2) = \sum_{k=0}^{n} \beta_{2k}{x_{2}^{k}}
	\end{equation}
where $\beta_{2k}$ is the $k$-th polynomial coefficient and $n$ represents its order. Although simple enough, one limitation with polynomials is that they are global functions of the input variable (as they are non-zero over an infinite region) which limits their use when the input space changes.

	\item Deterministic model 2: with Gaussian basis functions\\
	 \begin{align}
	 	\phi_{2k}(x_2) &= \text{exp}{\bigg(-\frac{(x_2 - b_k)^2}{c_k^2}\bigg)}\\
	 	f_2(x_2) &= \sum_{k=0}^{n} {a_k \, \text{exp}{\bigg(-\frac{(x_2 - b_k)^2}{c_k^2}\bigg)}}
	 \end{align}
	where $a_k$ is the amplitude, $b_k$ governs their location in the input space and $c_k$ controls their spatial width. Gaussian basis function is proposed as it allows better control with its location, width and scale parameter.
\end{itemize}

\subsection{Deterministic model validation}
The accuracy of the trained or identified deterministic model when it encounters the unseen data can be validated using $K$-fold cross-validation (CV). The idea is to divide the input set into $K$-subsets randomly. Then the metamodel is trained $K$-times and each time one of the subsets is left as a test set. On this test set, the training error measures are evaluated, for instance, the coefficient of determination, which is given by
 
\begin{equation}
		R^2 = 1-\frac{\sum_{i=1}^{n} (y_{i} - \hat{y}_i)^2}{ \sum_{i=1}^{n}(y_{i} - \bar{y})^2},
\end{equation}
where $y_i$ is the observed $i$-th datapoint, $\hat{y}$ is the predicted value and $\bar{y}$ is the average of the observed response. If the $R^2$-value for the $k$-th fitted subset is given by $R^2_k$, then the expected generalization error can be computed as follows \cite{friedman2017elements}:
\begin{equation}
	R^2_{CV} = \frac{1}{K}\sum_{k=1}^{K} R^2_k.
\end{equation}
To find the best value of $K$, the `bias-variance' trade-off needs to be considered. The value of $K$ depends on the size of the dataset \cite{friedman2017elements} and should be chosen such that the samples in the \textit{training} set and the \textit{test} set are representative of the larger dataset. A very common choice is to use $K = 5$ or $K = 10$ as they show lower variance \cite{breiman1992submodel, forrester2008engineering, friedman2017elements}. Another approach is to consider $K = N$, where $N$ is the size of the dataset. With this approach, each observed data-point is given an opportunity to be in the \textit{hold-out} test dataset. This approach, also referred to as \textit{leave-one-out} cross-validation, as one could imagine, is computationally expensive if the analysis is done on a huge dataset where a single model evaluation takes hours. Hence, $K$-fold CV (with $K=5$ and $K=10$) is the preferred method of deterministic model assessment in this study.

\subsection{Bayesian framework for uncertainty quantification (UQ)}\label{sec_Bayesian}
To quantify and better control the non-determinism in the responses, in the context of finite element analyses, a distinction has been made in the literature between \textit{possibilistic} (non-probabilistic) approaches and \textit{probabilistic} approaches \cite{faes_recent_2020}. Probabilistic metamodels and particularly, Bayesian approaches have garnered a lot of attention due to the flexibility they provide in choosing the plausible design alternatives based on prior-knowledge which imposes an inherent regularization. For instance, Bayesian framework was proposed for force reconstruction \cite{zhang2012bayesian}, inverse acoustic problems \cite{pereira2015empirical} and for design-driven validation approach \cite{chen2008design}.

In the Bayesian formulation, considering the GAM equation \eqref{eq:state_space_deterministic} and let $p(\btheta)$ denote the ``prior" probability density function (pdf) of the parameters $\btheta$, and let $p(\btheta|\boldsymbol{y})$ denote the ``full posterior distribution" of the parameters conditional on the observed (measured) data $\boldsymbol{y}$. Bayes' theorem provides an analytical relation to compute this posterior distribution \cite{bolstad2009understanding, gelman2013bayesian}:
\begin{equation}
	p(\btheta|\boldsymbol{y}) = \frac{f(\boldsymbol{y}|\btheta) p(\btheta)}{p(\boldsymbol{y})}.
\end{equation}
In the equation above,
\begin{itemize}
	\item $f(\boldsymbol{y}|\btheta) = \mathcal{L}(\btheta|\boldsymbol{y})$ is called the ``likelihood" of observing the data,
	\item the marginal distribution, $p(\boldsymbol{y}) = \int_{\btheta}f(\boldsymbol{y}|\btheta)p(\btheta)d\btheta$ is the ``evidence" of the observed data.
\end{itemize}
The aim here is to compute $p(\btheta|\boldsymbol{y})$, but as the evidence scales with the parameter space, the closed form of this integral is not always available which leads to its intractability. Nevertheless, only the shape of the posterior distribution is needed for sampling the parameters (up to a proportionality constant) \cite{bolstad2009understanding}. The shape of the un-normalized posterior density is given by:
\begin{equation}
	p(\btheta|\boldsymbol{y}) \propto f(\boldsymbol{y}|\btheta)p(\btheta).
\end{equation}
Stochastic simulation methods (Monte Carlo methods) are used to sample from this unscaled distribution \cite{bolstad2009understanding} which approximates the true distribution provided that the sample size is large enough. 

However, such algorithms remain inefficient for high-dimensional parameter space \cite{bolstad2009understanding}. Therefore, another class of methods called Markov Chain Monte Carlo (MCMC) methods \cite{hastie2017generalized} which is based on sequentially simulating draws from a Markov chain\footnote{A Markov chain is defined as a stochastic process where the conditional probabilities of the process at time $t$ given the states at all the previous times $(t-1,t-2,...,0)$ only depends on the single previous state at time $(t-1)$ \cite{bolstad2009understanding}.} are used in this study. The idea is to let the Markov chain run for long enough until it converges to a limiting (or stationary) distribution. Samples drawn after this initial run-in (also called ``burn-in") time are the draws from the approximated target posterior distribution \cite{bolstad2009understanding}. Metropolis-Hastings sampler and Gibbs sampler are typical examples of such methods but according to \cite{neal1993probabilistic}, these samplers can become extremely time-consuming as they explore the parameter space via inefficient random walks. 

In this study, No-U-Turn-Sampler (NUTS) is preferred, which is based on Hamiltonian Monte Carlo (HMC) (also called as Hybrid MC) due to its ability to adapt the tunable parameters of HMC i.e. step-size and the number of steps \cite{hoffman2014no}. 

\subsection{Parametric Bootstrapping} \label{sec_parametric_bootstrapping}
Bootstrapping is a data-based simulation method for statistical
inference which was introduced in 1979 \cite{efron1979bootstrap}. In principle, there are two types of Bootstrapping methods mentioned in the literature: \textit{parametric} (based on bootstrapping the input-output pairs) and \textit{non-parametric} which is based on bootstrapping the residuals \cite{efron_introduction_1993}. In this work, \textit{parametric} approach is used as the model has limited parameters that need to be estimated. Such method is useful when the estimator is a complex function of the parameters \cite{murphy_machine_2012}.

Let $\hat{\btheta}$ denote the statistical estimate of the parameter $\btheta$. Here, the bootstrap approach is used to assign measures of accuracy to such statistical estimates. The bootstrap algorithm works by randomly sampling $n$ values with replacement from the observed data ($\boldsymbol{y}$), also called as \textit{bootstrapped} samples $(\by^{*}_{\text{BS(1)}},.., \by^{*}_{\text{BS}(n)})$, followed by the evaluation of the corresponding bootstrap replications $\hat{\btheta}^{*}_{\text{BS(\textit{i})}}, \forall i = 1,2,..,n$. The
sample variances computed from these bootstrapped samples are the estimators of the variances of the parameters. The point estimates are obtained using the nonlinear least-squares approach, where the expected value $\mathbb{E}[\by|\bX,\be]$ is given by the deterministic models shown in \secref{sec_model_formulations}. Parametric bootstrap algorithm implemented in this work is as follows:
\begin{enumerate}
	\item get a point estimate $\hat{\btheta}$ from the observed data $\{\by,\bX,\be\}$ ($\by$ here can be the selected data points after considering the categorical variables, defined by the nuisance parameter $\be$)
	\item repeat for $i=1,...,B$
	\begin{enumerate}
		\item simulate variables $\bX^{*}_{(i)}$, $\be^{*}_{(i)}$ and $\eta_{(i)}$
		\item simulate data $\by^{\text{sim}}_{(i)}=f(\bX^{*}_{(i)}$, $\be^{*}_{(i)};\hat{\btheta}) + \eta_{(i)}$
		\item estimate $\btheta^{*}_{(i)}$ from $\{\by^{\text{sim}}_{(i)},\bX^{*}_{(i)}$, $\be^{*}_{(i)}\}$
	\end{enumerate}
	\item get an approximation of $p(\btheta|\by,\bX,\be)$ from the histogram of $\{\btheta^{*}_{(i)};i=1,...,B\}$
\end{enumerate}
\section{Application context: Interior cabin noise level assessments} \label{sec_application_context}

\subsection{Physical mechanisms}
The overall noise experienced inside the passenger cabin is a relative contribution of different sources coming from the engine, wheels, powertrain, and wind. At different vehicle speeds, some of these are more dominating than the others. Especially in electric vehicles (EVs), due to the absence of engine's masking noise, the tonal whine from electric motors is one of the most dominant sources of noise. Regardless of the type of propulsion, two background noise sources, which lie in the broadband frequency regime contributing towards the masking effect, are generally present during the real driving conditions:

\begin{enumerate}
	\item \underline{Vehicle aerodynamic noise}: At higher speeds around 100 kmph, aerodynamic noise remains a dominant source of noise and discomfort \cite{oettle_automotive_2017}. Three types of mechanisms associated to the aerodynamic noise can be found in the literature \cite{cerrato2009automotive, oettle_automotive_2017}. As shown in \figref{fig_aerodynamic_noise_generation_mechanism}, it is characterized by different types of sources that depend on the vehicle speed (or the flow speed) and the experimental conditions (tightly sealed vehicle, cross-wind, etc.). In this study, the flow speed considered is 140 kmph and 200 kmph, which is same as the conditions used during wind-tunnel testing. For such high speeds, Mach number, $M > 0.1$ and hence the dominant source of noise are the dipole types of sources where the sound intensity, $I_{\text{dipole}} \propto v^{6}$, where $v$ is the flow speed \cite{oettle_automotive_2017}. This prior-knowledge based on \textit{first principle} will be exploited to develop deterministic surrogate for aerodynamic noise. This paper is largely based on the analysis of this type of noise.
	\begin{figure}[h]
		\centering
		\begin{subfigure}{0.5\textwidth}
			\centering
			\includegraphics[scale=0.2]{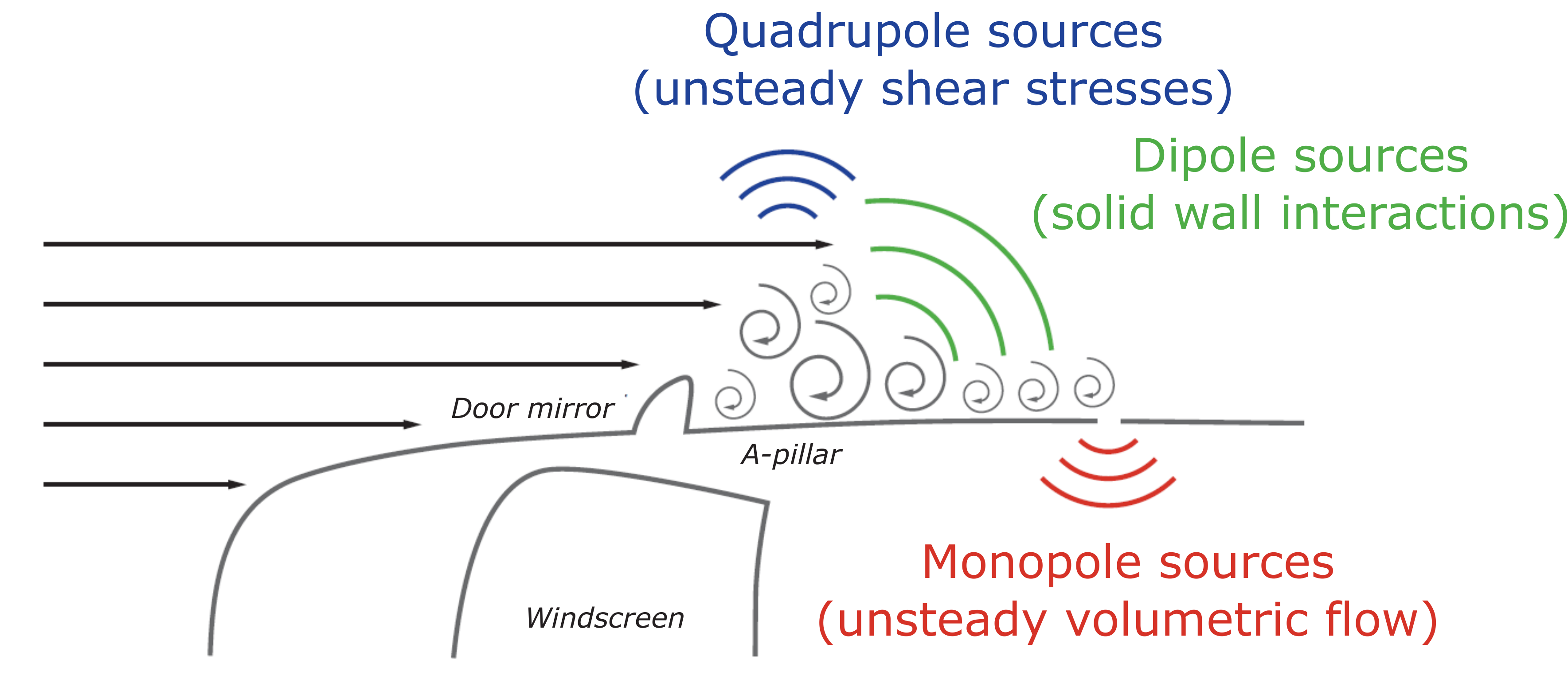} 
			\caption{Aerodynamic noise generation mechanism \cite{oettle_automotive_2017}}
			\label{fig_aerodynamic_noise_generation_mechanism}
		\end{subfigure}%
		\begin{subfigure}{0.5\textwidth}
			\centering
			\includegraphics[scale=0.2]{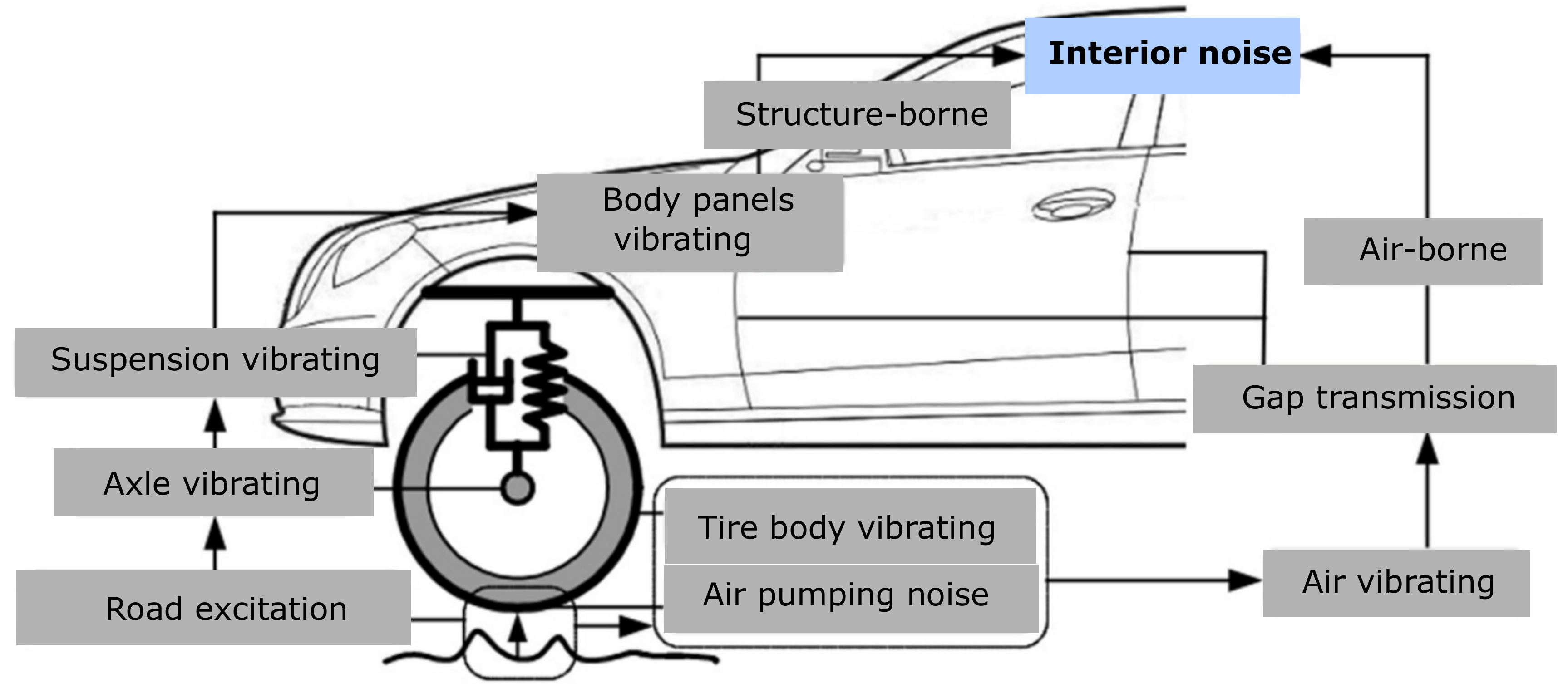} 
			\caption{Tire-road interaction phenomenon \cite{li2018literature}}
			\label{fig_tire_road_noise_image}
		\end{subfigure}
		\caption{Sources of vehicle masking noise due to wind (a) and due to tire-road interaction shown in (b)}
	\end{figure}
	
	\item \underline{Interior tire-road noise}: Physics behind the interior noise generated due to tire-road interaction is relatively more complex as it involves numerous design parameters. The structure-borne contribution (dominant at low frequencies $\approx 100$ Hz and low speeds) comes from the induced vibrations through sources such as tread impact and rolling deflections. Air-borne noise (dominant at higher frequencies $\approx 1$ kHz and higher speeds) is directly propagated through the medium due to air-pumping, air-turbulence and Helmholtz resonances \cite{li2018literature}. \figref{fig_tire_road_noise_image} shows various tire-road noise contributing factors.	
\end{enumerate}

\subsection{Model formulations} \label{sec_model_formulations}
The rest of the paper considers the following convention. $L$ is the sound pressure level (SPL) as a complex function of frequency $\bomega$, $\Ld$ is the SPL record available from database, $\Lp$ is the predicted SPL, $\bX$ the vector of predictor variables, such as operating conditions, $\btheta$ is the vector of parameters of a given sub-model, and $\eta$ denotes the fitting error, including both model errors and experimental noise.

Referring \eqref{eq:state_space_deterministic} and choosing the speed $\boldsymbol{v}$ and the frequency $\bomega$ to be the predictor variables,
\begin{align}
	\Ld &= L_{\text{aero}}+\eta\\\nonumber
	&= \bX\btheta + \boldsymbol{\eta}
\end{align}
where  $\bX = (\mathbf{1}, \boldsymbol{v}, \bomega)$ and $\btheta^T = (\alpha, \boldsymbol{\beta_{1}}, \boldsymbol{\beta_{2}})$.
Therefore, two deterministic nonlinear mappings $L_{\text{aero}}^{[1]}$ and $L_{\text{aero}}^{[2]}$ are formulated for aerodynamic wind noise considering $m=4$ and $n=6$ (the superscript in [$\,$] refers to two different models). As per \cite{mat_ali_wind_2018}, for dipole types of sources, $r=6$ is considered in the following equations,
\begin{equation}
	L_{\text{aero}}^{[1]} (\boldsymbol{v}, \bomega)= 10\, \text{log}_{10} \Bigg( \frac{b  \boldsymbol{v}^r}{C_{0}^{r-3}  10^{-12}} \Bigg) + \sum_{i=0}^{m}a_{i}\, \bomega^i; \label{eqn_deterministic_model_1}
\end{equation}
\begin{equation}
	L_{\text{aero}}^{[2]} (\boldsymbol{v}, \bomega)= 10 \, \text{log}_{10} \Bigg( \frac{b  \boldsymbol{v}^r}{C_{0}^{r-3}  10^{-12}} \Bigg) + \sum_{k=0}^{n} {a_k \, \text{exp}{\bigg(-\frac{(\bomega - b_k)^2}{c_k^2}\bigg)}}. \label{eqn_deterministic_model_2}
\end{equation}
Similarly, for tire-road noise, one physically informed model is:
\begin{equation}
	L_\text{tire} (\boldsymbol{v}, \bomega) = \bomega^{r1}  \sum_{i=0}^{m} a_i\,\boldsymbol{v}^i + 
	\boldsymbol{v}^{r2}  \sum_{k=0}^{n} {a_k \, \text{exp}{\bigg(-\frac{(\bomega - b_k)^2}{c_k^2}\bigg)}}. \label{eqn_tire_road_noise}
\end{equation}
\begin{figure}[!h]
	\centering
	\includegraphics[scale=0.4]{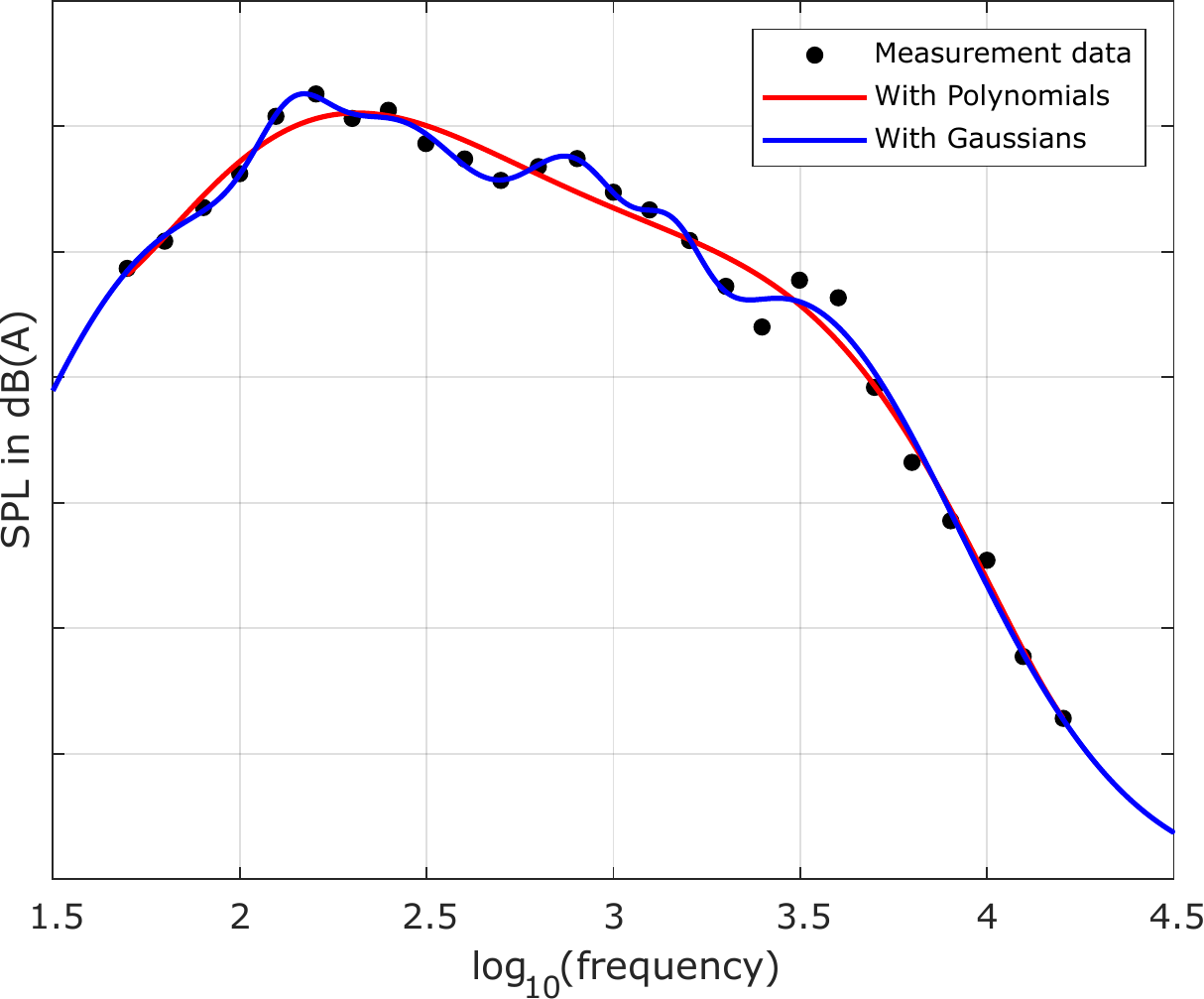} 
	\caption{Data fitting over the frequency range depending on the model choice. Note that the frequency is considered in its logarithmic form so as to ease the fitting process and be more interpretable.}
	\label{fig_Fitting_matlab}
\end{figure}

\subsection{$K$-fold Cross Validation}
For the analysis, two different vehicle body-types namely, Sedan and Hatchback are considered. The values of $K$, which is the number of parts the dataset is divided into, are chosen to be 5 and 10, with a total of 1000 runs carried out to analyze the model assessment process.
\begin{figure}[!h]
		\centering
	\begin{subfigure}{0.5\textwidth}
		\centering
		\includegraphics[scale=0.25]{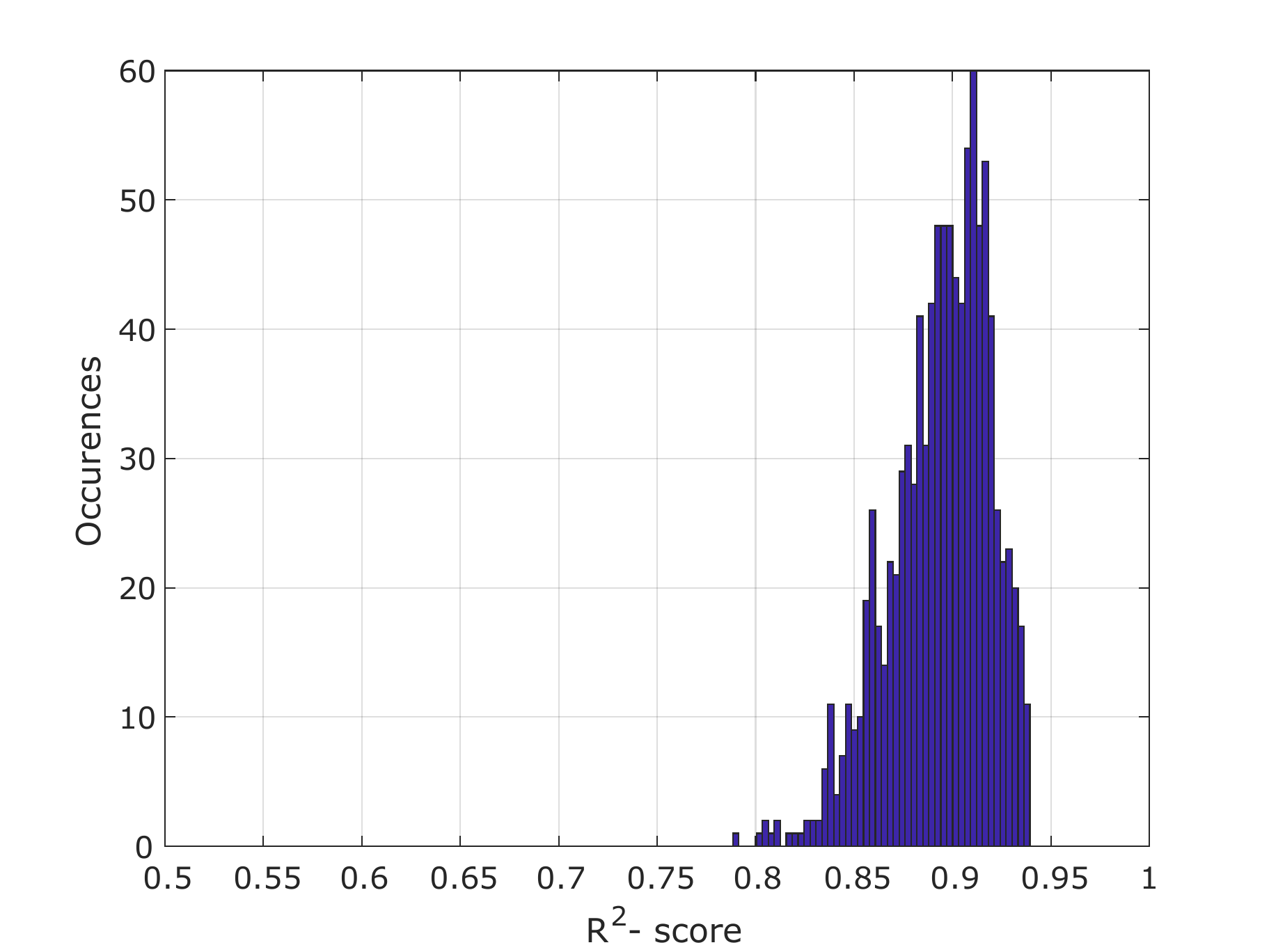} 
		\caption{$K$=5, vehicle body-type: Sedan}
		\label{fig_kCV_5_aeroModel_sedan}
	\end{subfigure}%
	\begin{subfigure}{0.5\textwidth}
		\centering
		\includegraphics[scale=0.25]{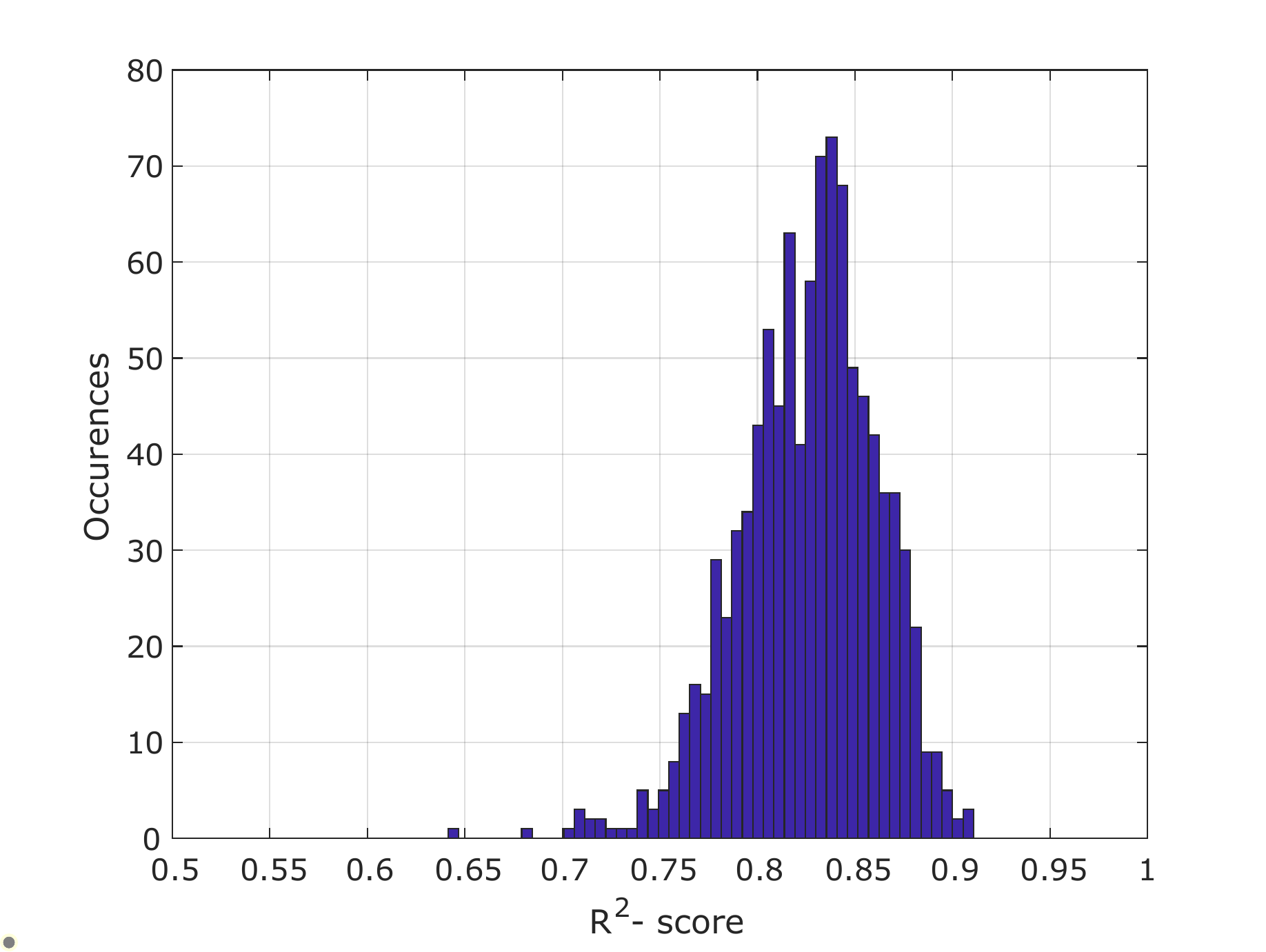} 
		\caption{$K$=10, vehicle body-type: Sedan}
		\label{fig_kCV_10_aeroModel_sedan}
	\end{subfigure}
	\caption{Histogram for $K$-fold CV for vehicle body-type Sedan}
\end{figure}
\begin{figure}[!h]
		\centering
	\begin{subfigure}{0.5\textwidth}
		\centering
		\includegraphics[scale=0.25]{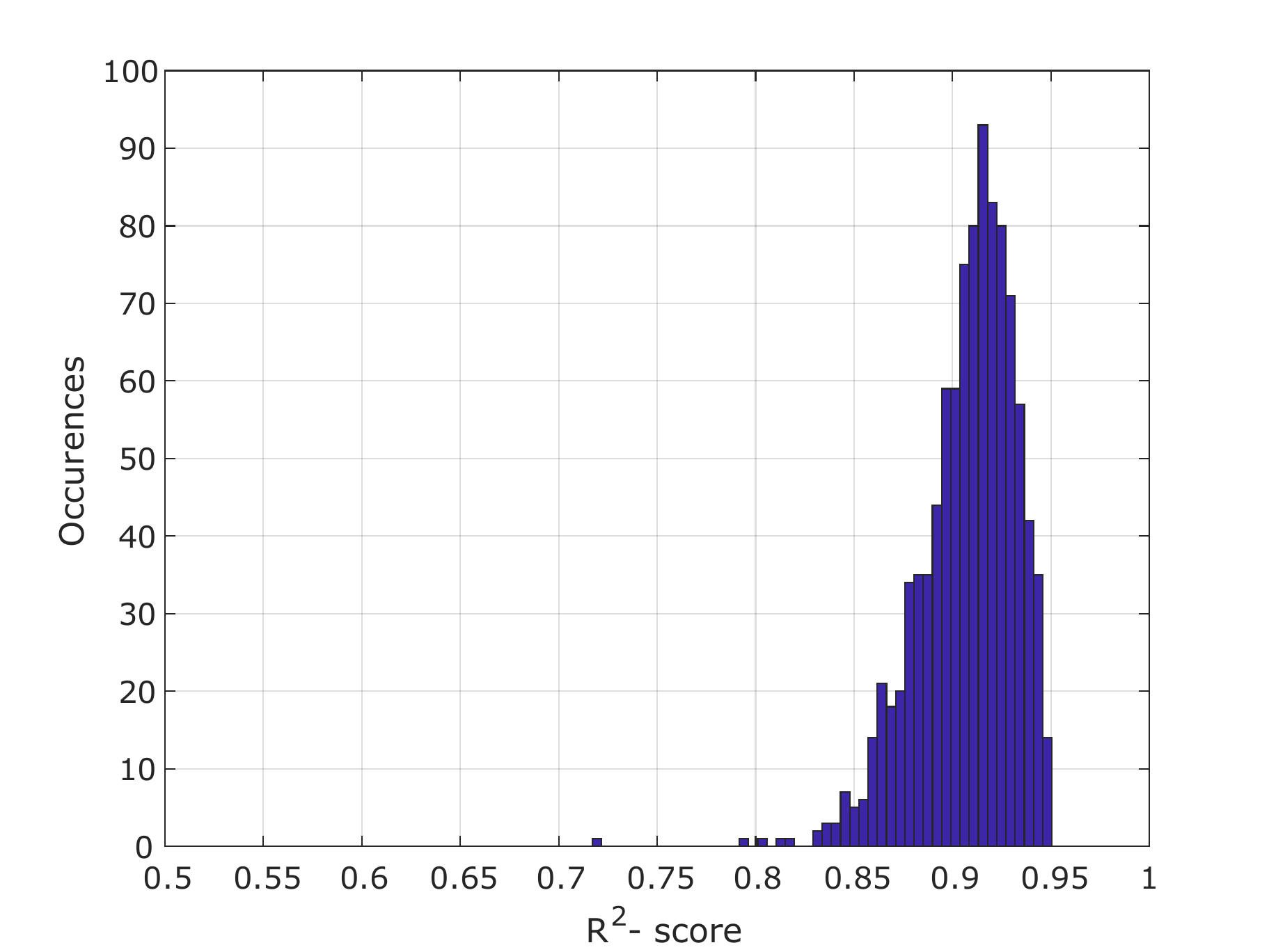} 
		\caption{$K$=5, vehicle body-type: Hatchback}
		\label{fig_kCV_5_aeroModel_hatchback}
	\end{subfigure}%
	\begin{subfigure}{0.5\textwidth}
		\centering
		\includegraphics[scale=0.25]{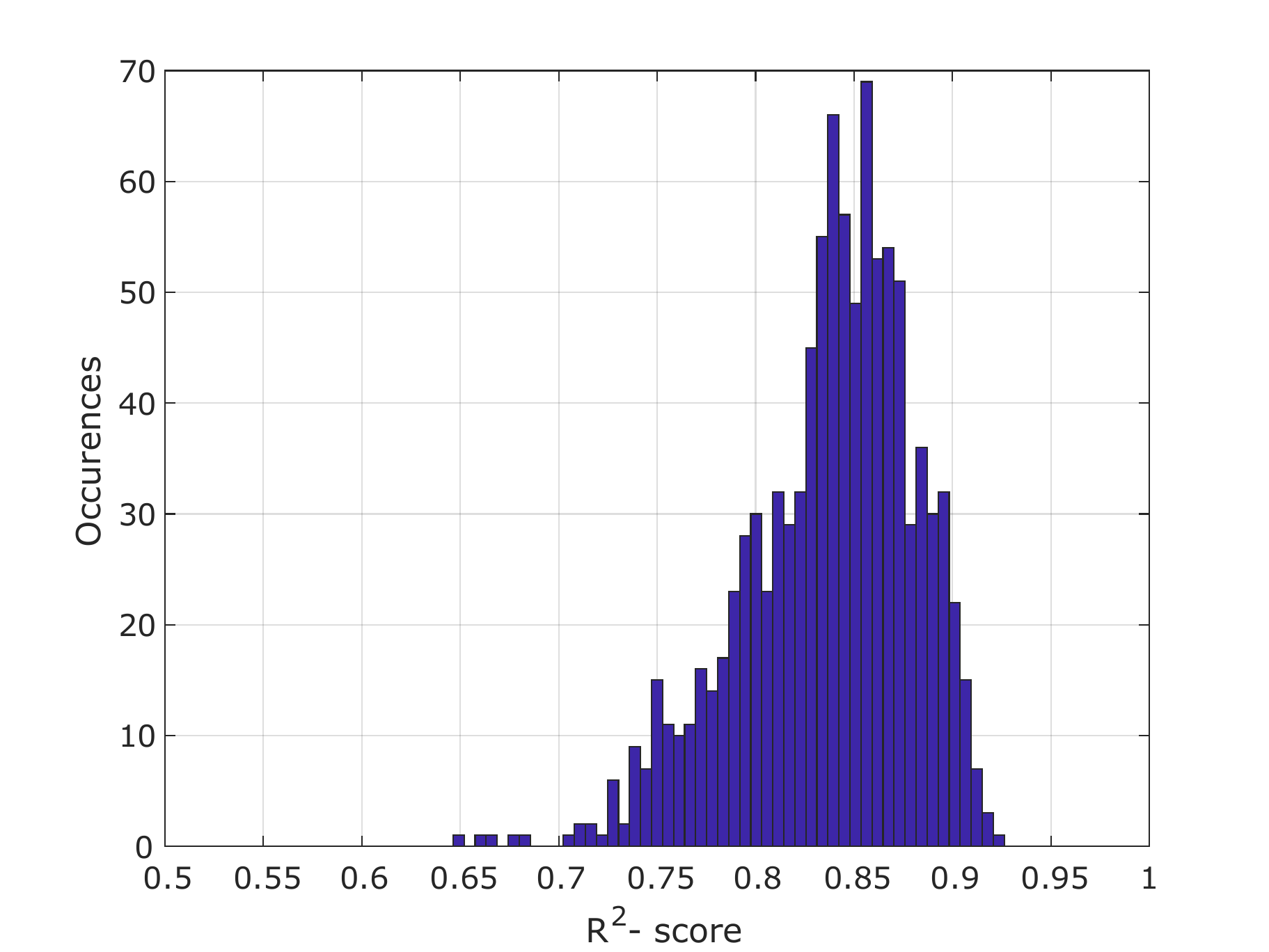} 
		\caption{$K$=10, vehicle body-type: Hatchback}
		\label{fig_kCV_10_aeroModel_hatchback}
	\end{subfigure}
	\caption{Histogram for $K$-fold CV for vehicle body-type Hatchback}
	\label{fig_Histogram_kCV_vehicle_body_type_Sedan_Hatchback}
\end{figure}
\begin{table}[!htbp]
	\centering
	\begin{tabular}{*5l}
		\toprule
		{} &  \multicolumn{2}{c}{\textit{Sedan}} & \multicolumn{2}{c}{\textit{Hatchback}}\\
		\midrule
		{}   & $K = 5$   & $K = 10$    & $K = 5$   & $K = 10$\\
		$\mu$   &  0.89 & 0.83   & 0.91  & 0.84\\
		$\sigma^2$   &  6.39e-04 &  0.0012   & 6.08e-04  & 0.0017\\
		\bottomrule
	\end{tabular}
	\caption{Comparison of mean and variance of $R^2_{CV}$-values for two different vehicle body types}
	\label{tab_comparison_of_kCV}
\end{table}
Histograms of the $K$-fold CV analysis for two different vehicle body types with polynomial basis functions can be seen in \figref{fig_Histogram_kCV_vehicle_body_type_Sedan_Hatchback} and summary of its mean and variance is shown in \tabref{tab_comparison_of_kCV}. It is observed that if the  training dataset is divided into 5 parts, then the CV accuracy is $\approx 90\%$ with a very small variance.

\subsection{Data selection for UQ} \label{sec_data_selection_for_UQ}
\begin{figure}[!htp]
	\centering
	\begin{subfigure}{0.5\textwidth}
		\centering
		\includegraphics[scale=0.4]{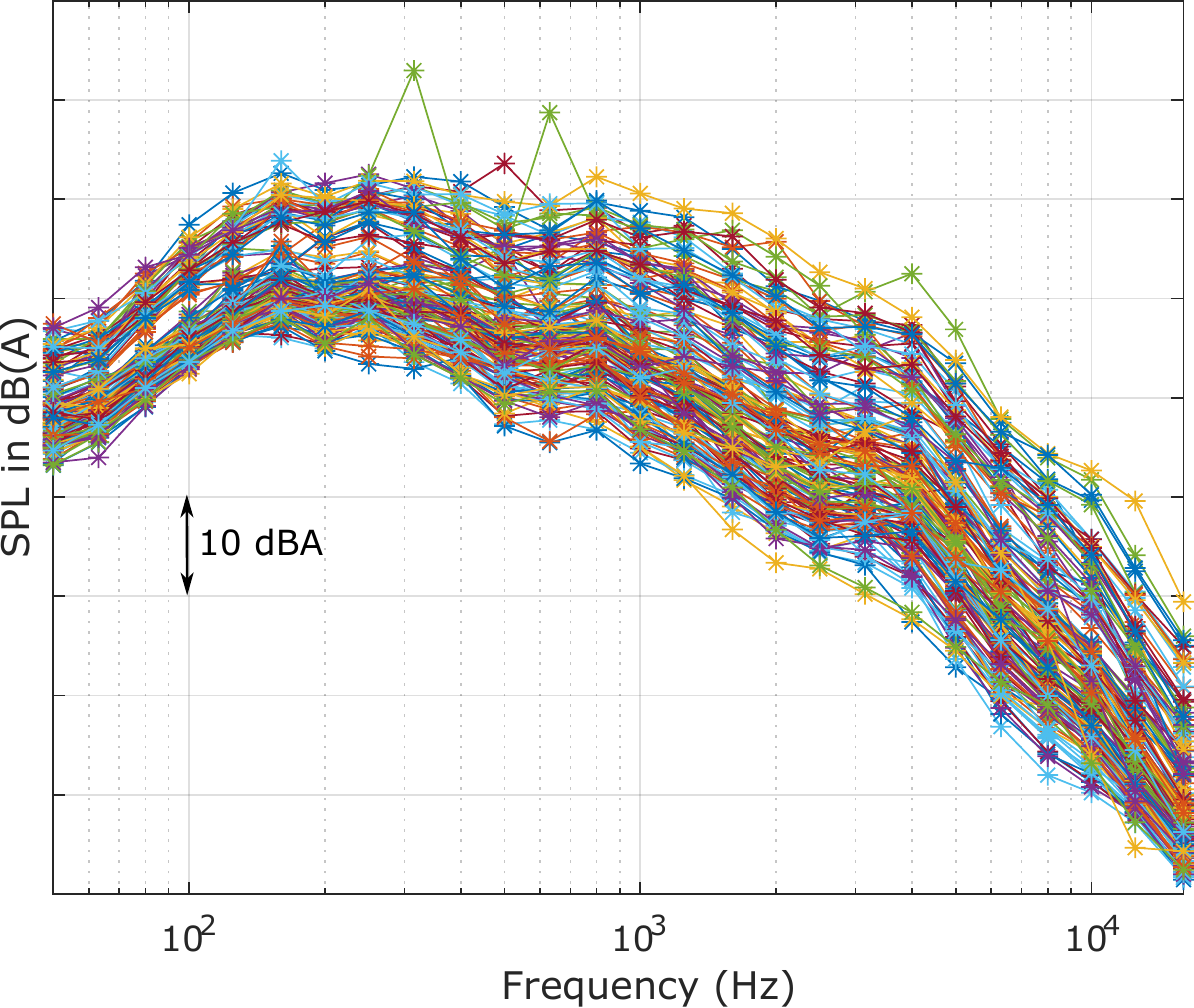} 
		\caption{Entire measurement database}
		\label{fig_SPL_aero_data_measurement_onlyForSedan_showingLargeVariance_NoYticks}
	\end{subfigure}%
	\begin{subfigure}{0.5\textwidth}
		\centering
		\includegraphics[scale=0.4]{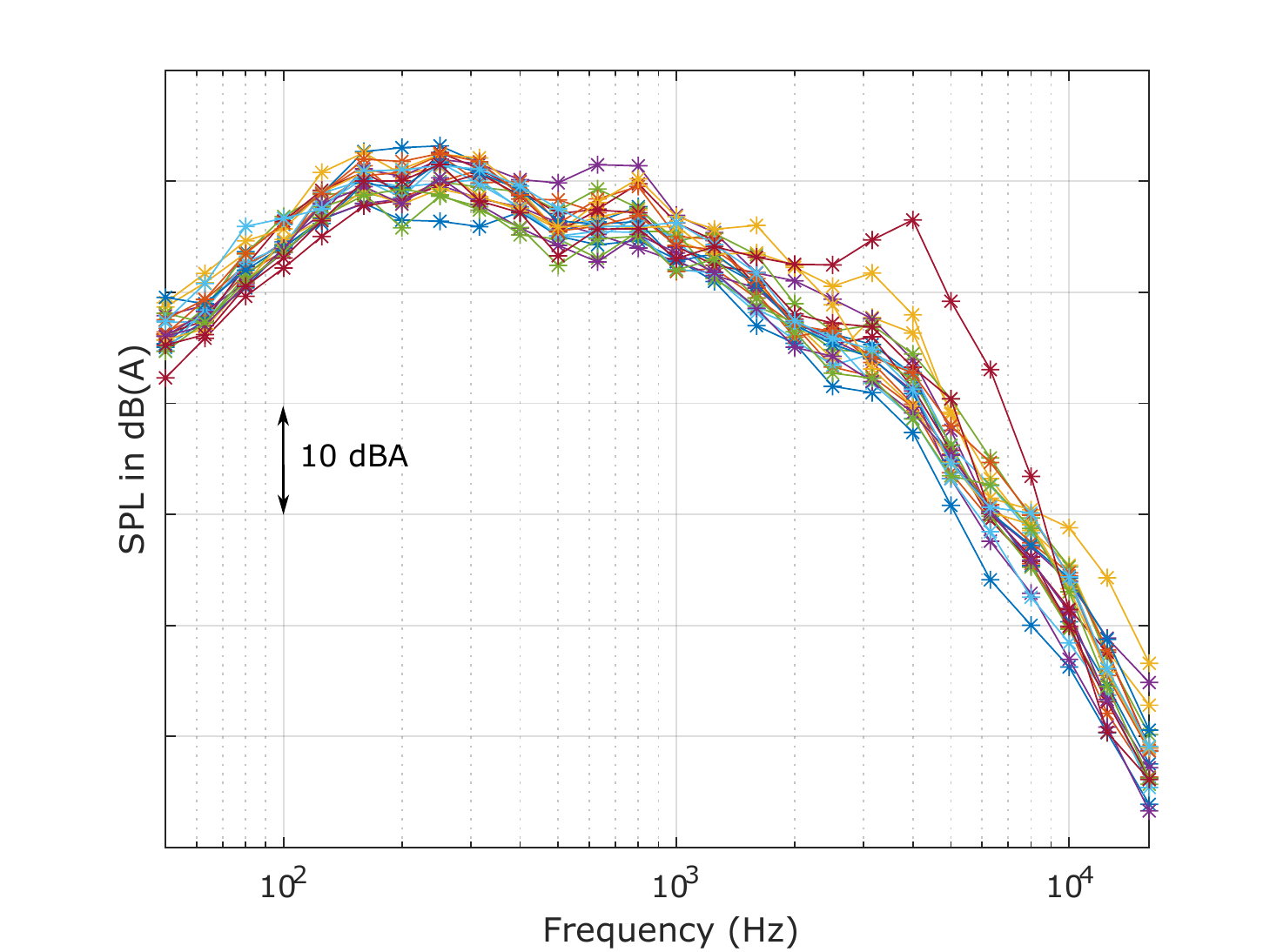} 
		\caption{Database selected for probabilistic modelling}
		\label{fig_SPL_aero_data_measurement_withCategoricalVar_NoYticks}
	\end{subfigure}
	\caption{Measurement data showing inherent variability due to different types of vehicles and operating conditions}
\end{figure}

An important aspect of stochastic modelling is to be able to capture the inherent variability in the data. For instance, large dispersion in the measured database can be seen in \figref{fig_SPL_aero_data_measurement_onlyForSedan_showingLargeVariance_NoYticks} which corresponds to a large number of vehicles tested, depending on operating conditions. This being stated, real engineering practice is to refine this database and select the data corresponding to the specific category of vehicle under consideration. Therefore, \textit{categorical} variables are defined, that are used as selection parameters in the database. For instance, for aerodynamic wind-noise, the database is filtered as per one particular: 1) body design, 2) segment, 3) energy, 4) target market, roof-type, 5) state, and 6) measurement position. \figref{fig_SPL_aero_data_measurement_withCategoricalVar_NoYticks} shows the refined database on the basis of categorical variables which is used for the Bayesian learning approach in this paper. This particular choice of categorical variables will define the model $\mathcal{M}_1$.

\subsection{Bayesian hierarchical models for UQ in \normalfont{$ L_{\text{aero}}$}}
In this section, the two models, given by \eqref{eqn_deterministic_model_1} and \eqref{eqn_deterministic_model_2}, are formulated in the Bayesian context. It is assumed that the observed data in the measurement set, given by $\Ld$ in dB, is distributed according to Normal distribution, which belongs to the exponential family of distribution as mentioned in \secref{sec_GAM}, with mean $\mu$ and variance $\sigma_y^2$. Also, all the unknown parameters (non-categorical) are modelled as random variables with a prior probability distribution. Choice of prior distributions here are supposed to be in line with the objective knowledge the analyst has on the parameters before observing the data (through domain-expertise, physical laws, etc.). If too little objective information is available, non-informative priors can be chosen. Also, for the sake of simplicity, priors that are conjugate to the likelihood are preferred, refer \cite{brogna2018probabilistic, gelman2013bayesian,bolstad2009understanding}.
\begin{figure}[htp]
	\centering
	\begin{subfigure}{0.5\textwidth}
		\centering
		\includegraphics[scale=0.4]{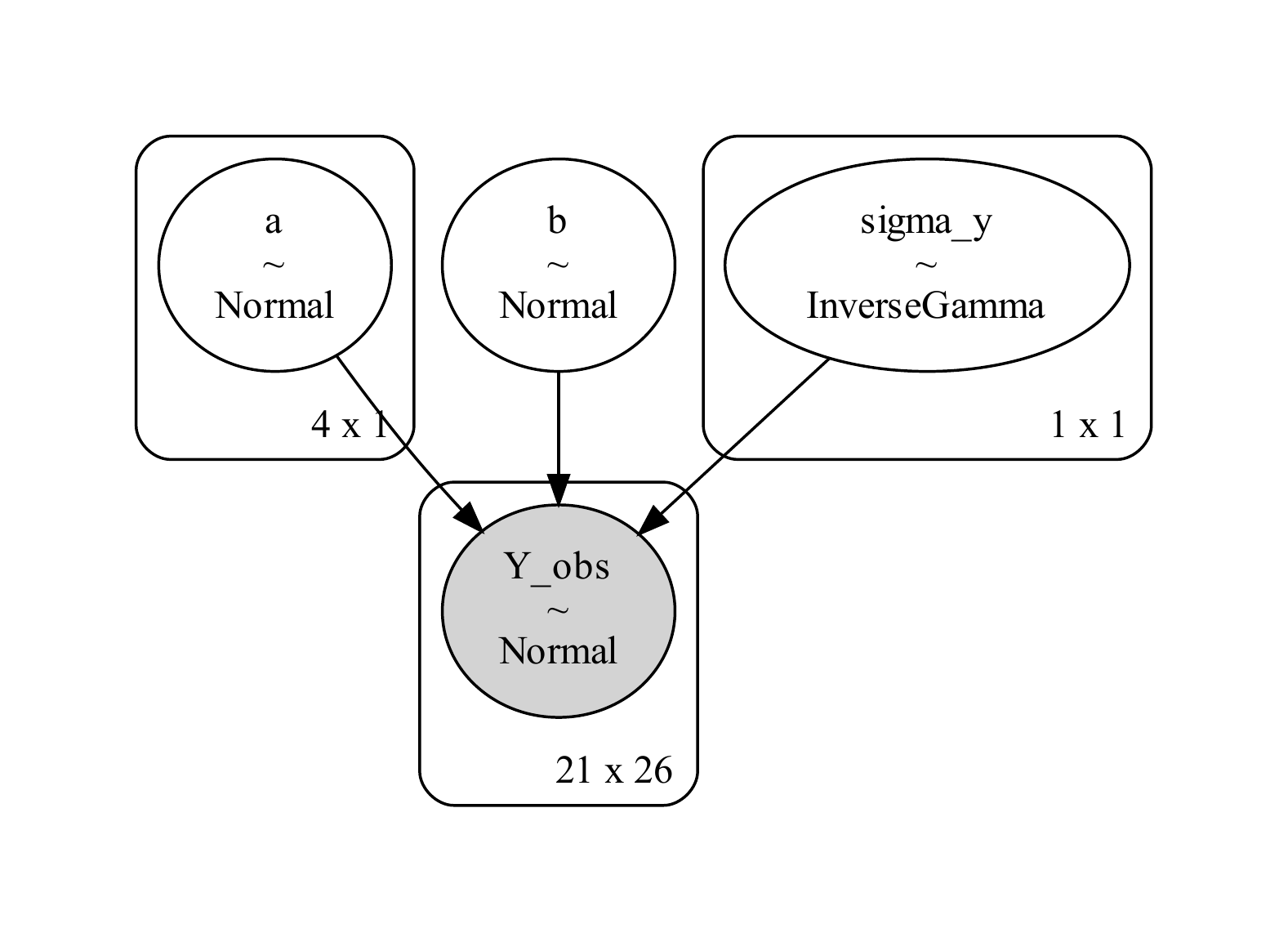} 
	\end{subfigure}%
	\begin{subfigure}{0.5\textwidth}
		\centering
		\includegraphics[scale=0.4]{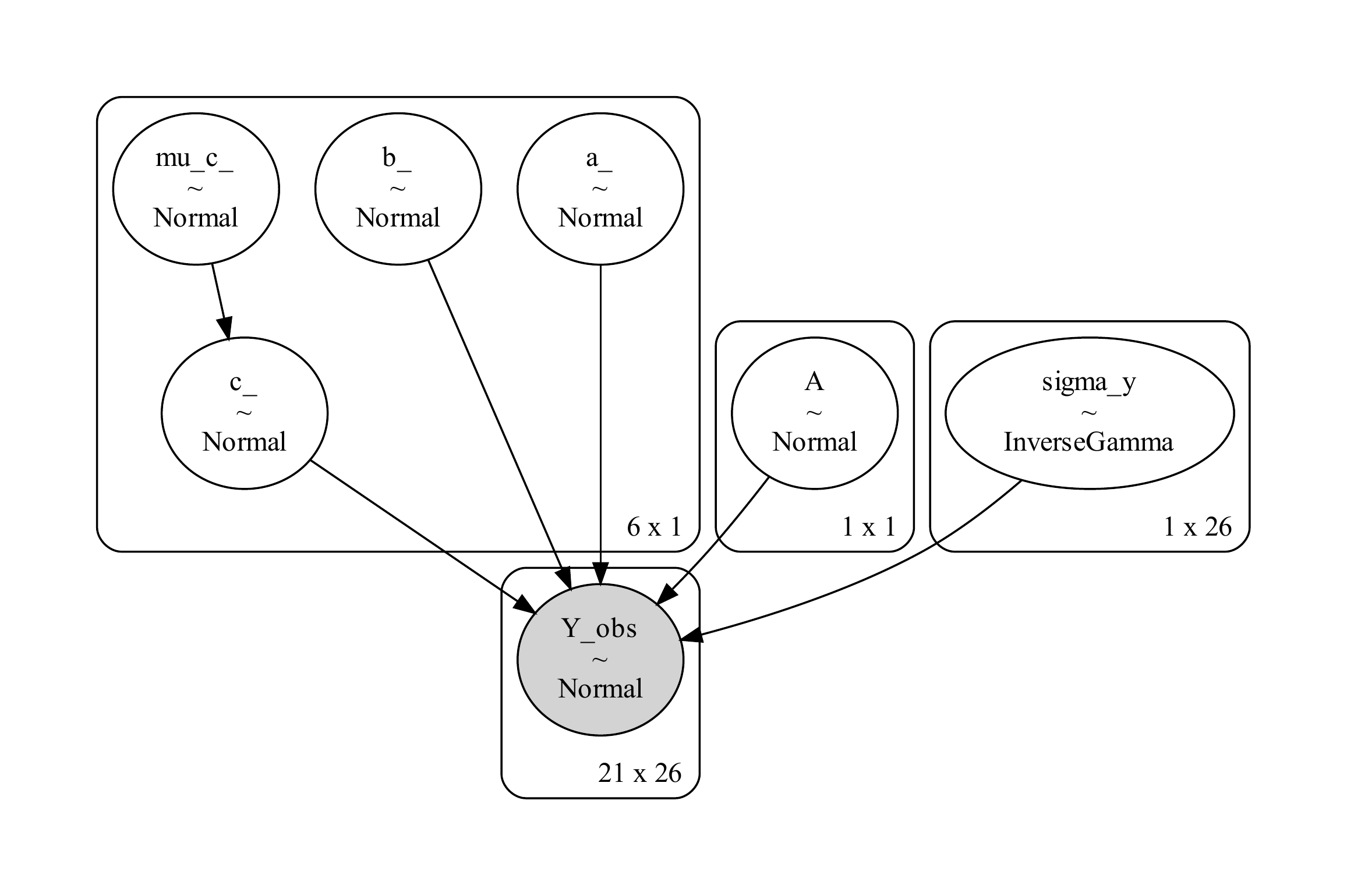} 
	\end{subfigure}
	\caption{Bayesian Hierarchical models: (left) first model with polynomial basis, and (right) second model with Gaussian basis}
	\label{fig_Bayesian_Hierarchical_models}
\end{figure}

\figref{fig_Bayesian_Hierarchical_models} provides a way to represent full joint probability of the random variables in the form of a graph using the assumption of conditional independence. Let us say that there are two events $A$ and $B$, then they are conditionally independent given $Y$, if and only if we can express $p(A,B | Y) = p(A|Y)\, p(B|Y)$. These graphical models display the independence relationships and are defined in terms of directed acyclic graphs. The nodes in the graph are modelled as random variables and edges encode their relationships \cite{jordan_graphical_2004}. Also, the plate notation is exploited for capturing the replication in graphical models, which indicates that the data is independently and identically distributed (i.i.d) \cite{murphy_machine_2012}. The observed or deterministic variables (here, the measurement data represented as the likelihood function) are shown as shaded region and the unobserved random variables are represented as unshaded circular regions. The arrows on the edges, for example $a \rightarrow Y_{\text{obs}}$, indicate that $a$ is causing or influencing $Y_{\text{obs}}$. Moreover, the hyperparameters are not influenced by any other parameter in the hierarchy. Each distribution is characterized by its own hyperparameters, which are deterministic values and are chosen by the analyst, for example, a Normal distribution can have two hyper-parameters: $\mu$ for location and $\sigma^2$ for scaling. The higher we move up the hierarchical model, lesser is the impact of that parameter on the likelihood of observing the data and on the inference. If there's not much information available about the hyperparameters, then non-informative \textit{hyperpriors} can be assigned to them. In \figref{fig_Bayesian_Hierarchical_models}, $\mu_c$ is the hyperprior for the random variable $c$ and is characterized by its own hyperparameters $(\mu_{cc}, \sigma_{cc})$ \cite{gelman2013bayesian}. For the sake of visual clarity, we have not shown hyper-parameters in these graphs.

Considering $N$ to be the total number of measurement data available and $n_\theta$ be the total number of parameters to learn, the two Bayesian models are formulated as
\begin{itemize}
	\item Bayesian Model 1 (BM1) with polynomial basis functions
	\begin{equation}
		\begin{aligned}
			\Ld| \bX,\btheta, \eta &\sim \mathcal{N} (L_{\text{aero}}^{[1]} (\boldsymbol{v}, \bomega), \sigma^2_y)\\
			\sigma^2_y &\sim \text{InvGamma} (\alpha_y, \beta_y)\\
			\theta_i &\sim \mathcal{N}(\mu_i, \sigma_i^2), \forall i=1,..,n_{\theta}\\
		\end{aligned}
	\label{eq_BM1}
	\end{equation}
	
	\item Bayesian Model 2 (BM2) with Gaussian basis functions with heteroscedasticity
	\begin{equation}
		\begin{aligned}
			\Ld| \bX,\btheta, \eta &\sim \mathcal{N} (L_{\text{aero}}^{[2]} (\boldsymbol{v}, \bomega), \boldsymbol{\sigma}^2_y)\\
			\boldsymbol{\sigma}^2_y &\sim \text{InvGamma} (\boldsymbol{\alpha}_y, \boldsymbol{\beta}_y),\\
			\alpha &\sim \mathcal{N} (\mu_\alpha, \sigma_\alpha)\\
			\boldsymbol{a} &\sim \mathcal{N}(\mu_a, \sigma_a)\\
			\boldsymbol{b} &\sim \mathcal{N}(\alpha_b, \beta_b)\\
			\boldsymbol{c} &\sim \mathcal{N}(\mu_c, \sigma_c)\\
			\boldsymbol{\mu}_c &\sim \mathcal{N}(\mu_{cc}, \sigma_{cc})\\
		\end{aligned}
	\label{eq_BM2}
	\end{equation}
\end{itemize}

\section{Model convergence and results comparison} \label{sec_model_convergence}
Bayesian computational statistics allows us to approximate the posterior distribution and solve complex Bayesian models which could have been, otherwise, mathematically intractable. In order to make sure that the samples drawn are independently and identically distributed, several visual and numerical diagnostic tools have been developed, for relevant metrics refer \cite{martin2022bayesian}.

The two Bayesian models formulated in \eqref{eq_BM1} and \eqref{eq_BM2} are specified using an open-source library PyMC3 \cite{salvatier2016probabilistic}, through which NUTS sampler can be used that automatically adapts the step-size parameter and is efficient in generating independent samples \cite{hoffman2014no}. For each model, 4 chains were simulated with 10,000 samples in each. The burn-in samples were set to 2,000 to let the chain converge to its stationary distribution. 
\begin{figure}[h]
	\centering
	\begin{subfigure}{0.5\textwidth}
		\centering
		\includegraphics[scale=0.32]{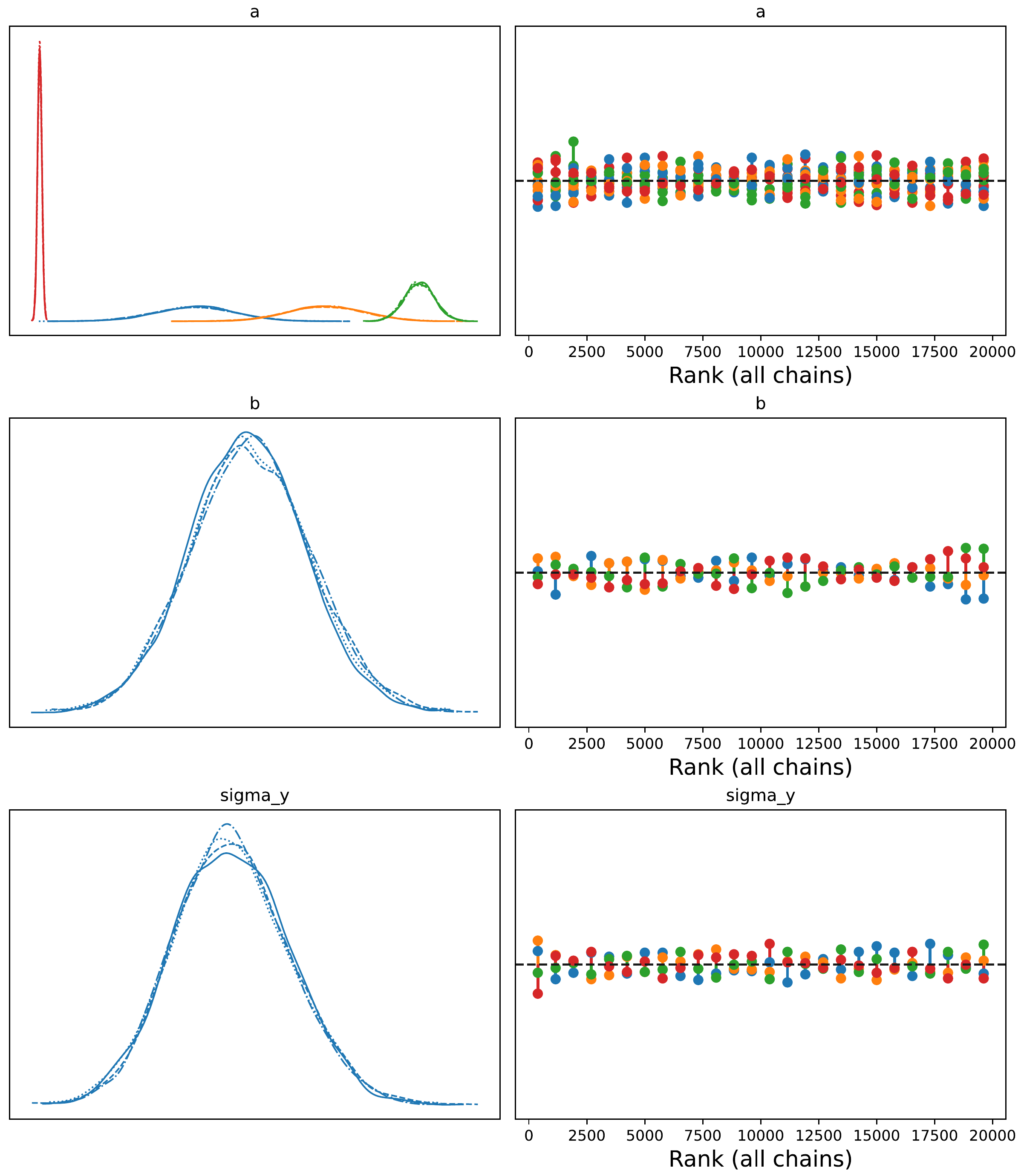} 
		\caption{Convergence results for BM1}
		\label{fig_aero_polynomials_Traceplots_4chains}
	\end{subfigure}%
	\begin{subfigure}{0.5\textwidth}
		\centering
		\includegraphics[scale=0.35]{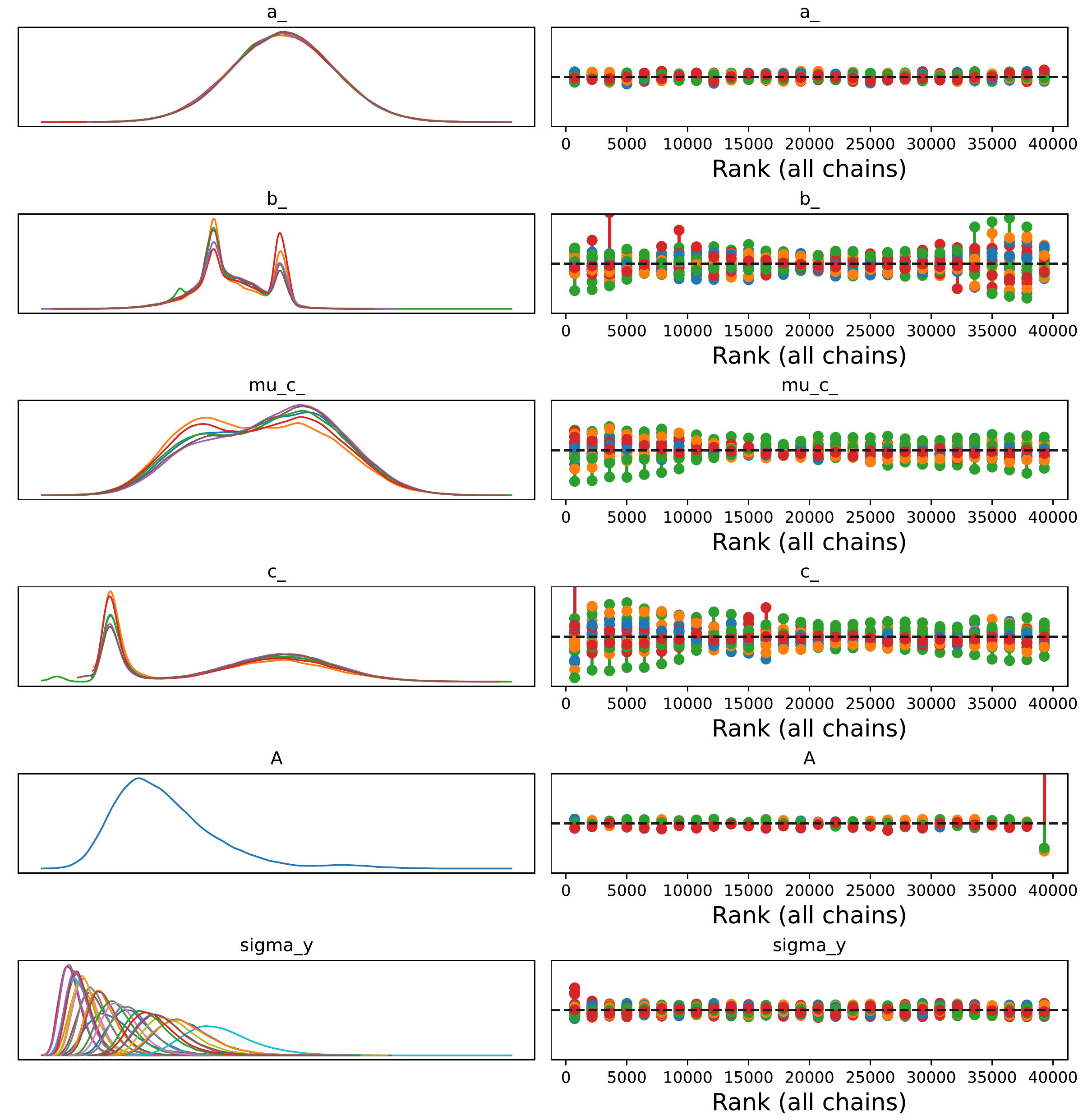} 
		\caption{Convergence results for BM2}
		\label{fig_aero_gaussians_Traceplots_4chains}
	\end{subfigure}
	\caption{MCMC convergence results for both the Bayesian models simulated with 4 chains and 10,000 samples each. Left columns of both the sub-figures are the kernel density estimate (KDE) plot corresponding to each unobserved random variable and the right columns consist of the rank plots}
	\label{fig_mcmc_convergence_statistics}
\end{figure}
The convergence statistics are shown in \figref{fig_mcmc_convergence_statistics}. It shows that the marginalized posterior distribution of the unknown parameters in the model. Moreover, rank plots can be observed, which are the histograms of the ranked posterior samples plotted separately for each chain \cite{vehtari2021rank}. If all chains are drawing from the same posterior (i.e. there's good mixing), the ranks in each chain should be uniform. Chains in BM1 show good mixing behavior as compared to the BM2 where the rank plots for some parameters deviate from uniformity. This is due to the fact that there exists a bimodal marginal posterior distribution for some parameters.

A very commonly used numerical diagnostic is the potential scale reduction factor or Gelman-Rubin statistic denoted by $\hat{R}$. It is computed for a particular parameter $\theta$ as the ratio of the standard deviation of all the samples of $\theta$ from all chains, to the root mean square of the individual within-chain standard deviations \cite{gelman2013bayesian}. 


It is expressed as
\begin{equation}
	\hat{R} = \sqrt{\frac{\hat{\text{var}}^+ (\theta|\Ld)}{W}},
\end{equation}  
where $W$ denotes within-chain variances and $\hat{\text{var}}^+ (\theta|y)$ is the marginal posterior variance of the parameter given by
\begin{equation}
	\hat{\text{var}}^+ (\theta|\Ld) = \frac{N-1}{N} W + \frac{1}{N}B
\end{equation}
where $N$ is the total number of draws per chain and $B$ is the between-chain variance (refer \cite{gelman2013bayesian, vehtari2021rank} for detailed formulations). Ideally, $\hat{R} \approx 1.0$ as the variance between the chains should be same as the variance within-chain.

For BM1, $\hat{R} = 1.0$ for all the parameters in the model which indicates that the chains are mixed well and the draws are from the same posterior distribution. For BM2, the number of parameters are significantly higher than model 1 and $1.0 \leq \hat{R} \leq 1.06$ which again denotes that the chains converged well and the samples are not much correlated.
\begin{figure}[h]
	\centering
	\begin{subfigure}{0.5\textwidth}
		\centering
		\includegraphics[scale=0.4]{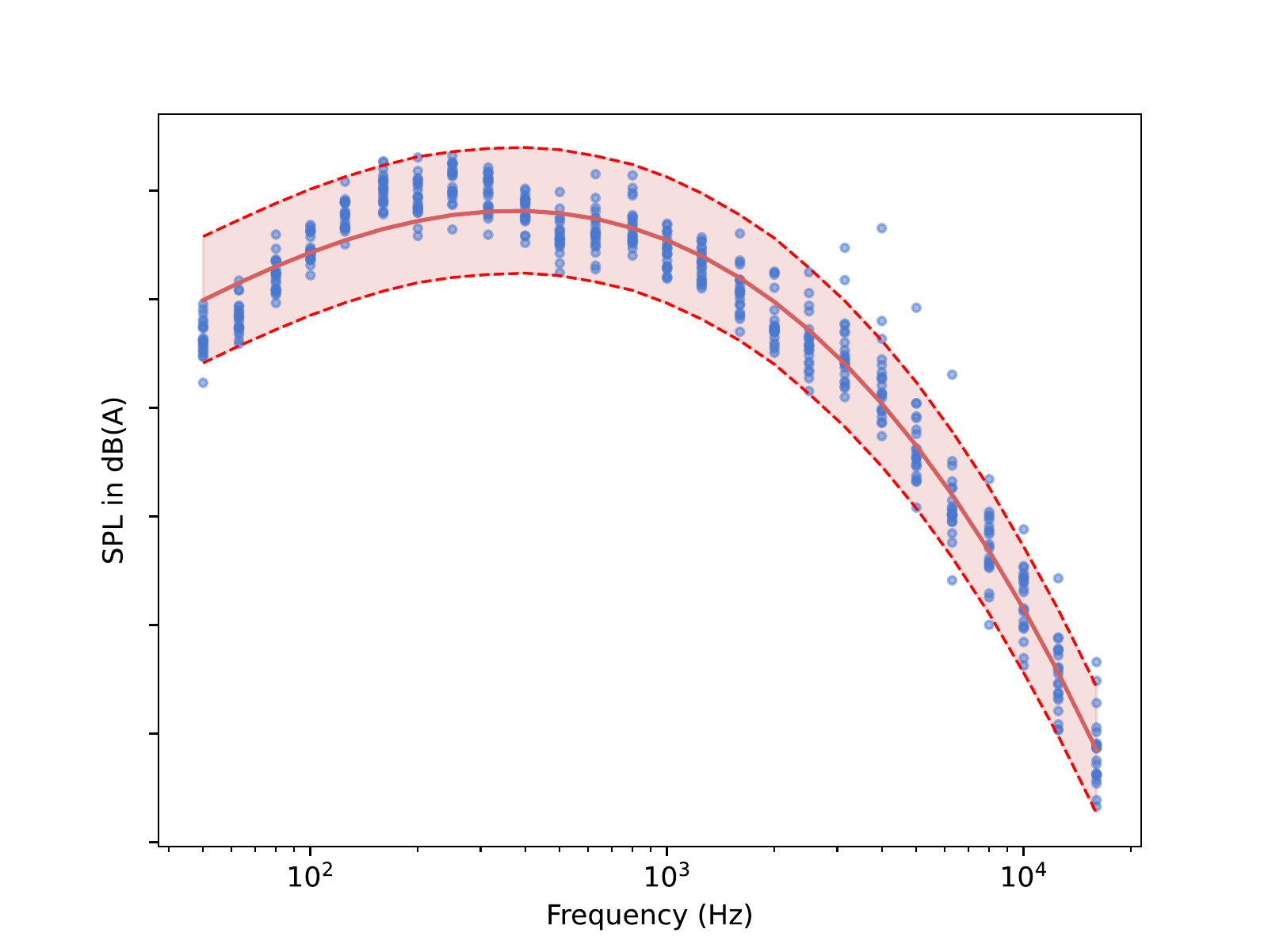} 
		\caption{BM1}
		\label{fig_Aero_Polynomials_NoYticks}
	\end{subfigure}%
	\begin{subfigure}{0.5\textwidth}
		\centering
		\includegraphics[scale=0.4]{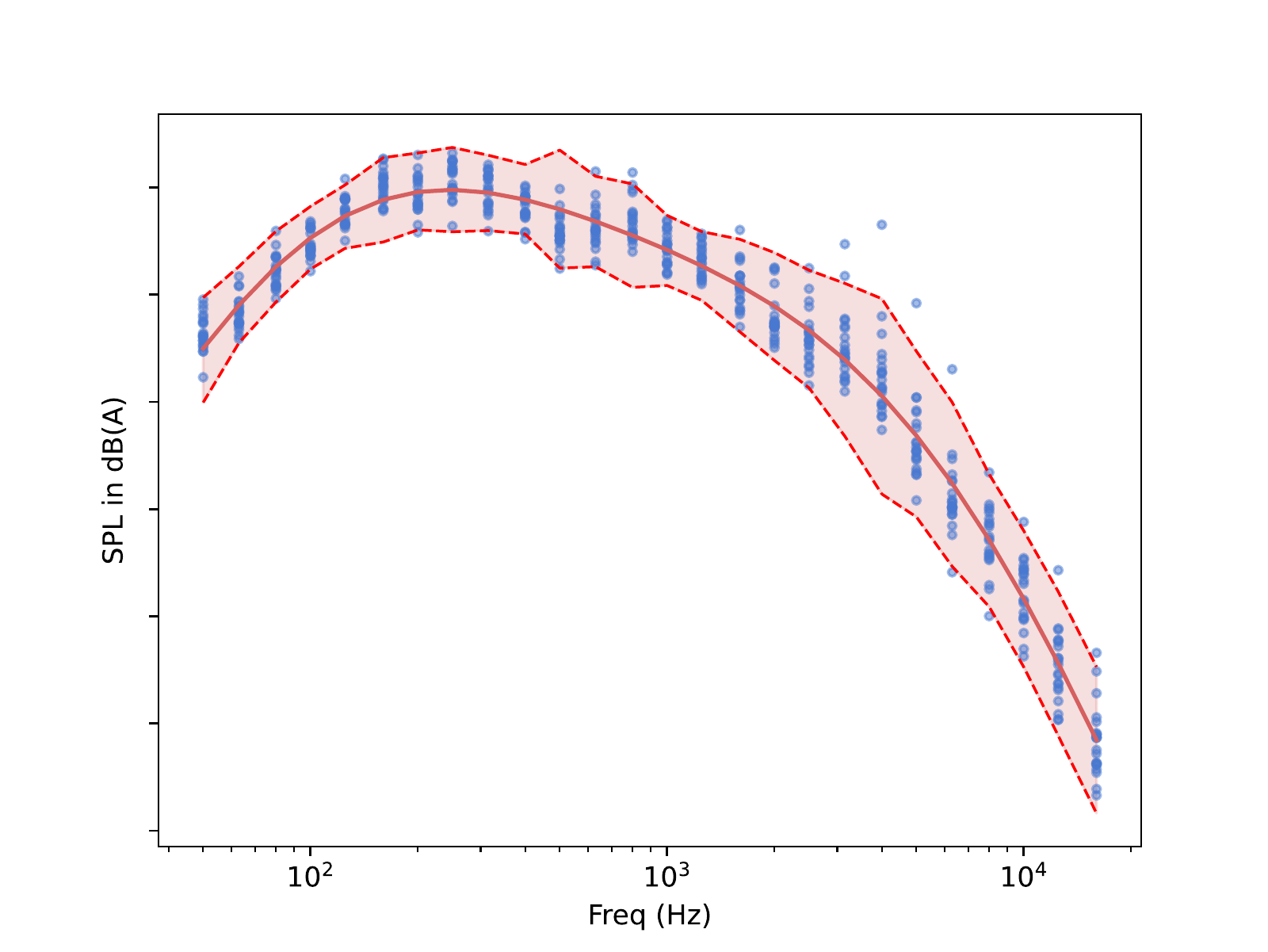} 
		\caption{BM2}
		\label{fig_Aero_heteroscedasti_NoYticks}
	\end{subfigure}
	\caption{Posterior predictive distribution for the two Bayesian models. \textcolor{blue}{Blue} dots refer to the measurement data, \textcolor{red}{red} solid line is the mean of the samples drawn from the posterior predictive distribution, and \textcolor{red}{red} dashed lines along with the shaded region represent the $95\%$ Bayesian credible interval}
	\label{fig_posterior_pred_distribution}
\end{figure}

The posterior predictive distribution, which is the distribution of the potential or future data $p(\Lp|\btheta,\bX,\be)$ is according to the posterior distribution of the parameters $p(\btheta|\Ld,\bX,\be)$. This is the model's prediction after seeing the observed data $\Ld$. It is given by
\begin{align}
	p(\Lp|\Ld,\bX,\be) =& \int f(\Lp|\btheta,\bX,\be)~p(\btheta|\Ld,\bX,\be)d\btheta \nonumber\\
	=& \int \mathcal{L}(\btheta;\Lp,\bX,\be)~p(\btheta|\Ld,\bX,\be)d\btheta
\end{align}
where $\be$ is the vector of nuisance variables that characterizes a particular model, for instance, model $\mathcal{M}_1$ described in \secref{sec_data_selection_for_UQ}.

The posterior predictive distribution plots are shown in \figref{fig_posterior_pred_distribution}. It can be seen that BM1, which has significantly less number of parameters, do not capture some intricate peaks in the model. On the other hand, with BM2, the heteroscedasticity allows the model to capture physical phenomenon quite well. Indeed, this comes at a cost of higher number of parameters and it requires more time to draw samples from the joint distribution of the parameters. Depending on the final use of the metamodel, BM1 or BM2 will be preferred.

The developed models should not be too primitive that they miss or fail to capture the valuable information in the data nor too complex that they fit the noise in the data. Therefore, computing the out-of-sample predictive accuracy for the fitted Bayesian models becomes essential. This is also called as generalization error, which is a measure of how well the model performs when it sees the data not used to fit it. Cross-validation and information criteria are the two methods for estimating this out-of-sample predictive accuracy. In this work, Bayesian Leave-One-Out cross validation (LOO-CV) \cite{vehtari2017practical} is used, where the hold-out dataset consists of a single data-point. The expected log pointwise predictive density (ELPD) is given by,
\begin{equation}
	\text{ELPD}_{\text{LOO-CV}} = \sum_{i=1}^{n} \log\int f(\Ld_i|\btheta,\bX,\be) ~p(\btheta|\Ld_{-i},\bX,\be)d\btheta \label{eq_ELPD}
\end{equation}
where $\Ld_i$ represents the $i$-th datapoint and $\Ld_{-i}$ indicates all datapoints except $i$-th. However, computing \eqref{eq_ELPD} is costly since $\btheta$ is not known a priori. To circumvent this, \eqref{eq_ELPD} is approximated from a single fit using Pareto smoothed importance sampling LOO-CV (PSIS-LOO-CV)\cite{vehtari2017practical}. The idea is that the observations $\Ld$ are assumed to be conditionally independent so that \eqref{eq_ELPD} is approximated using normalized weights $w$ as
\begin{equation}
	\text{ELPD}_{\text{PSIS-LOO}} = \sum_{i}^{n} \log \sum_{j}^{s} w_i^j~f(\Ld_i|\btheta^j,\bX,\be).
\end{equation}
Weights $w$ are computed using importance sampling where
\begin{equation}
	w_i^j = \frac{p(\btheta^j | \Ld_{-i})}{p(\btheta^j|\Ld)} \propto \frac{1}{f(\Ld_i|\btheta^j,\bX,\be)}.
\end{equation}
To avoid overshooting the variance of importance weights, a smoothing procedure is applied using the generalized Pareto distribution. The estimated $\hat{k}$ parameter of the Pareto distribution allows to detect highly influencial observations. These are the observations that have a significant effect on the posterior distribution when they are considered in the hold-out set. For well specified data and models, the value of $\hat{k}$ remains low ($\approx 0.5$), according to \cite{vehtari2017practical,gabry2019visualization}.

\begin{figure}[h]
	\centering
	\begin{subfigure}{0.5\textwidth}
		\centering
		\includegraphics[scale=0.4]{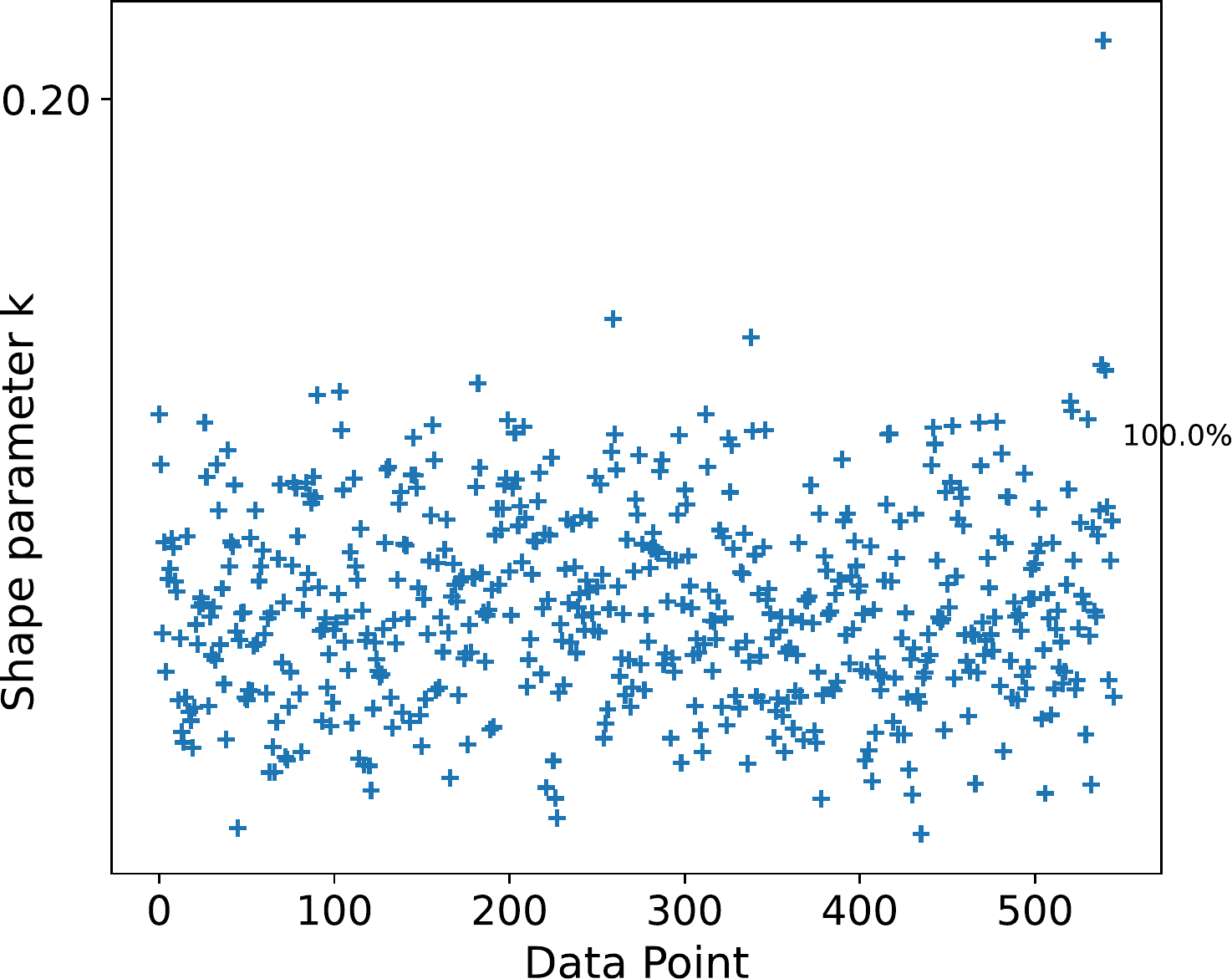} 
		\caption{BM1}
		\label{fig_LOO_CV_khat_aeroPoly}
	\end{subfigure}%
	\begin{subfigure}{0.5\textwidth}
		\centering
		\includegraphics[scale=0.4]{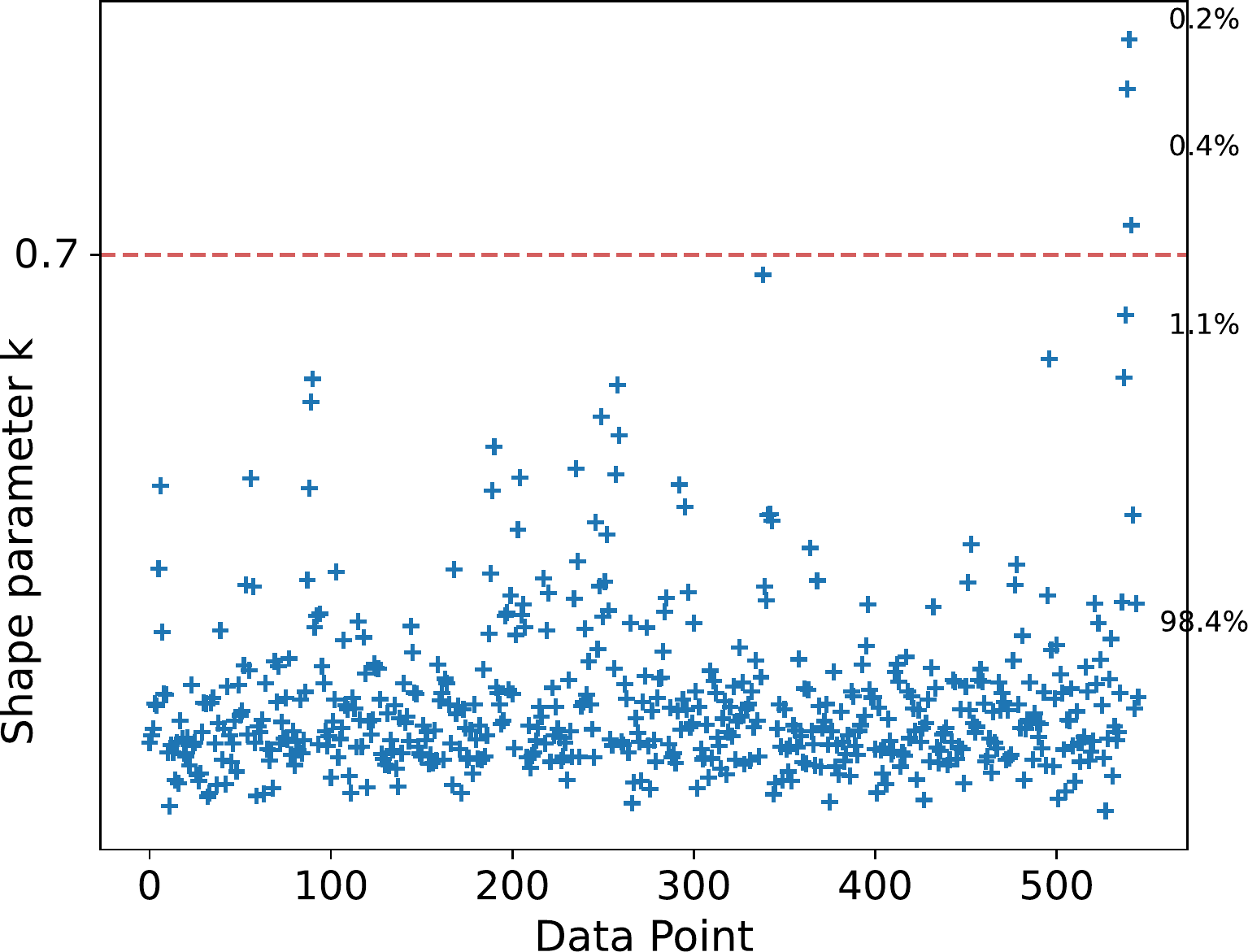} 
		\caption{BM2}
		\label{fig_LOO_CV_khat_aeroGaussians}
	\end{subfigure}
	\caption{The $\hat{k}$ diagnostics from PSIS-LOO-CV}
	\label{fig_LOO_CV_khat}
\end{figure}

\tabref{tab_comparison_of_LOO} summarizes the LOO-CV approach for the two Bayesian models. Rank denotes the ranking for the two models on the basis of LOO value. Penalization term gives an indication of the total number of effective parameters in the model and standard error is the error of LOO computations where a lower value is preferred \cite{vehtari2017practical}. Clearly from \tabref{tab_comparison_of_LOO}, BM2 with Gaussian basis function and heteroscedasticity is preferred over BM1 due to its lower error and better LOO value. Moreover, from \figref{fig_LOO_CV_khat}, we notice that $\hat{k}$ for all the datapoints in BM1 have a very low value ($<0.3$), whereas for BM2, there are 0.4\% of the data points that lie above the threshold 0.7. These are the same points at 4 kHz, 5 kHz, and 6.3 kHz, that the model is not able to explain due to their outlier behaviour.   
\begin{table}[!htbp]
	\centering
	\begin{tabular}{*5c}
		\toprule
		{} &  Rank & LOO & penalization-LOO & Standard error\\
		\midrule
		Gaussians   &  0& -1259.2   & 40.7  & 22.6\\
		Polynomials   & 1 &  -1367.6   & 5.2  & 24.1\\
		\bottomrule
	\end{tabular}
	\caption{Comparison of PSIS-LOO-CV for the Bayesian models 1 and 2}
	\label{tab_comparison_of_LOO}
\end{table}

\begin{figure}[h]
	\centering
	\begin{subfigure}{0.5\textwidth}
		\centering
		\includegraphics[scale=0.3]{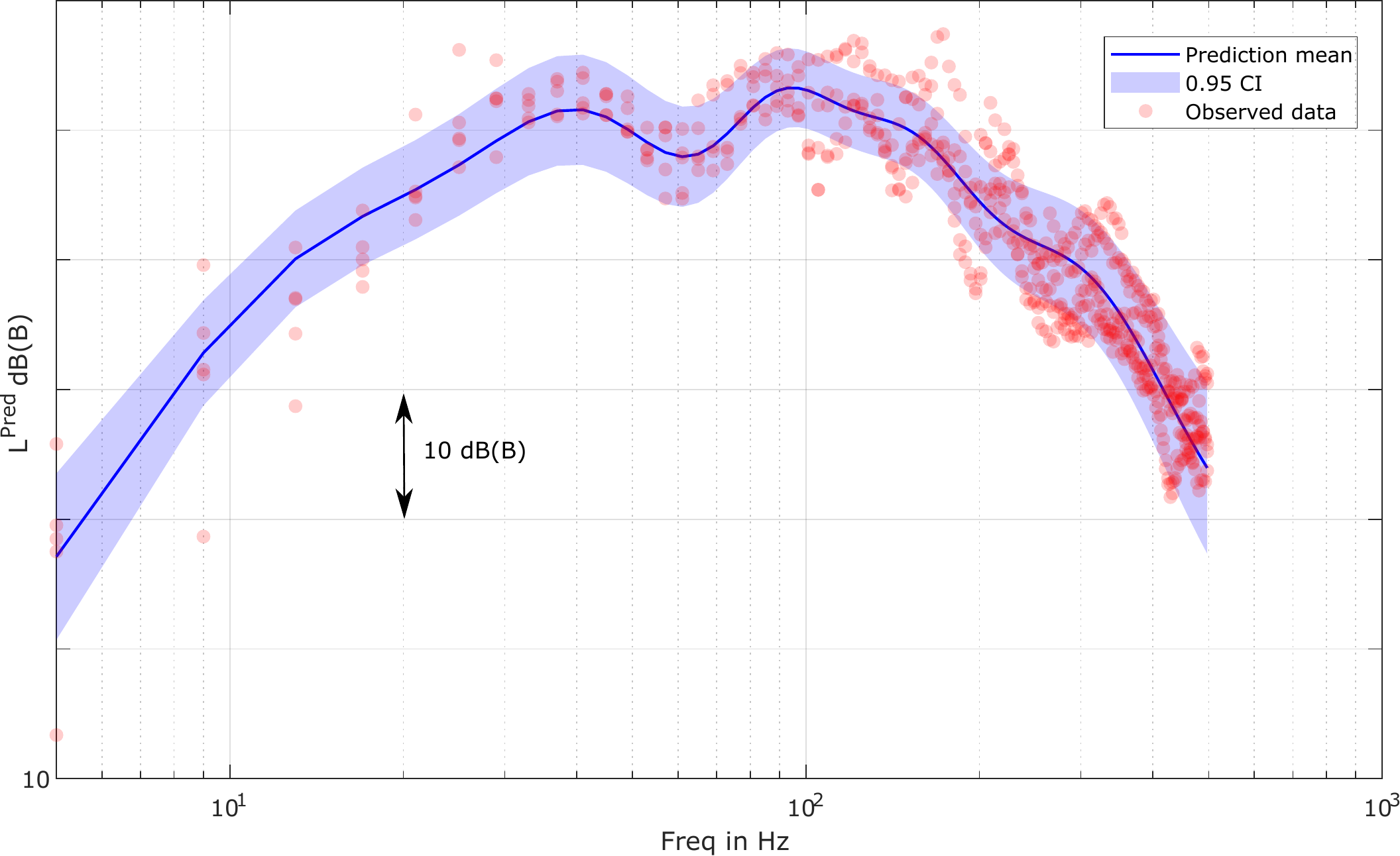} 
		\caption{SPL Prediction on multiple test data}
		\label{fig_tire_Bootstrapping_SPL_prediction_logScaleCorrected}
	\end{subfigure}%
	\begin{subfigure}{0.5\textwidth}
		\centering
		\includegraphics[scale=0.3]{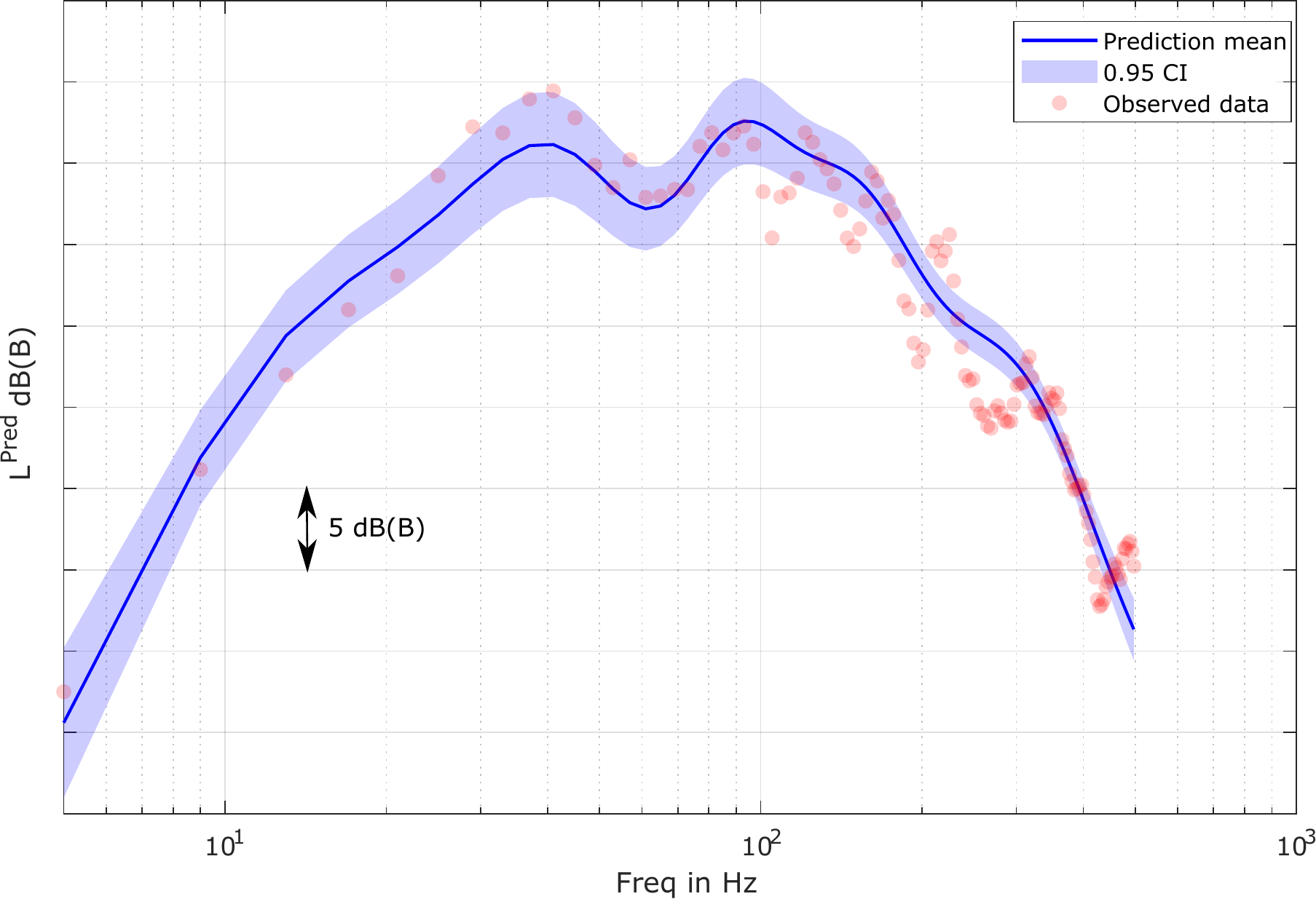} 
		\caption{SPL Prediction on a single test data}
		\label{fig_tire_Bootstrapping_SPL_prediction_categoricalVars_at_speed50_logScaleCorrected}
	\end{subfigure}
	\caption{Predictions from the parametric bootstrap approach. \textcolor{red}{Red} dots are the unseen test data, prediction mean is shown in solid \textcolor{blue}{blue} line with shaded region being the 95\% confidence interval}
	\label{fig_parametric_bootstrap}
\end{figure}

\begin{figure}[h]
\centering
\includegraphics[scale=0.3]{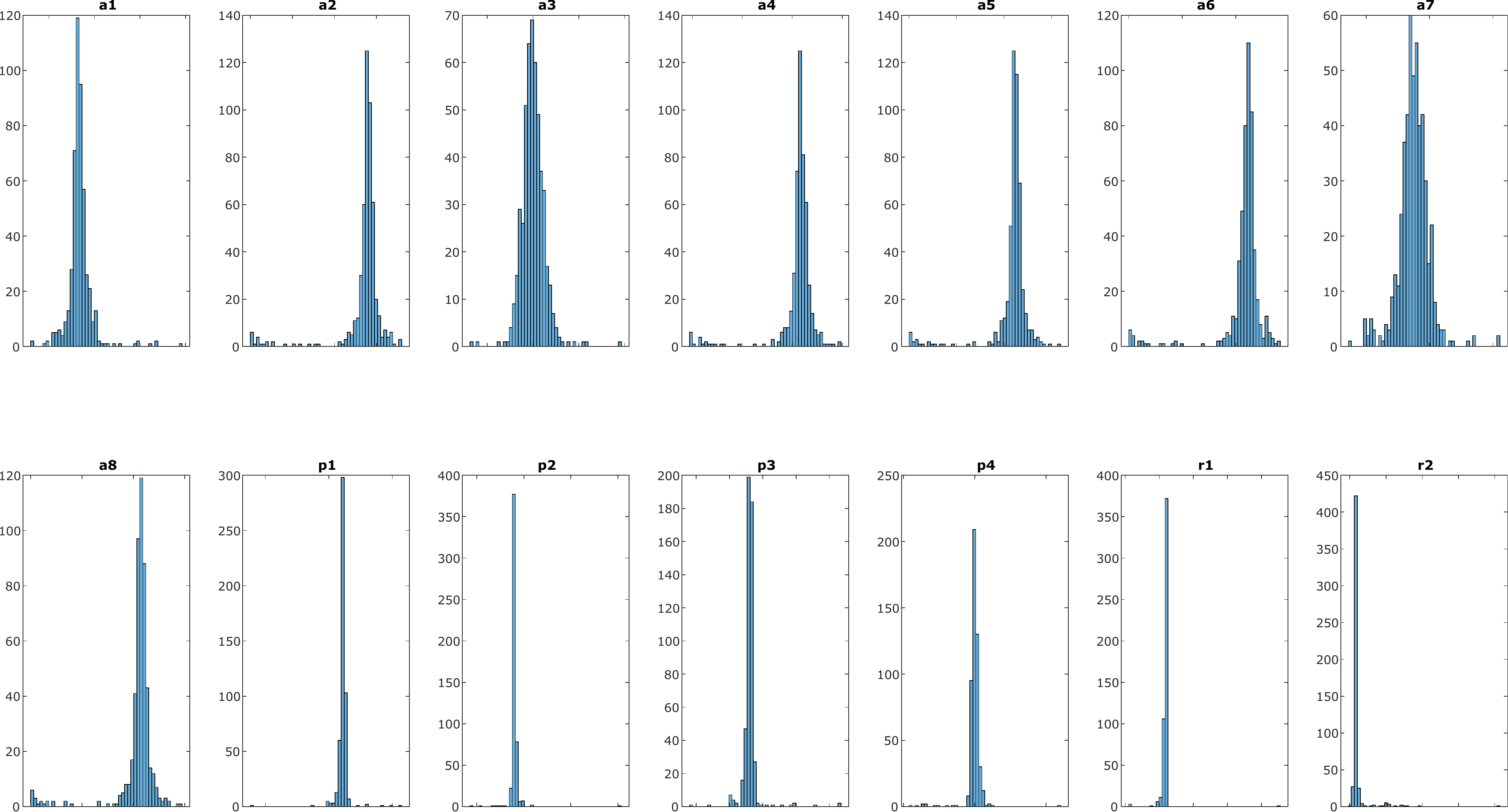} 
\caption{Histograms of the parameters estimated through bootstrapping}
\label{fig_tire_Bootstrapping_parameter_Histogram}
\end{figure}

For interior tire-road noise, parametric bootstrapping algorithm was implemented since the likelihood function given by \eqref{eqn_tire_road_noise} is accurate only up to 84\% (as per the $R^2$-value). The parametric bootstrap algorithm is explained in \secref{sec_parametric_bootstrapping} which considers \eqref{eqn_tire_road_noise} to generate simulated data from the point estimates. \figref{fig_parametric_bootstrap} shows the prediction on randomly test data set for multiple inputs (for eg, at speeds 50 kmph, 70 kmph, 90 kmph) and for one particular test speed. The histograms for the estimated parameters can also be seen in \figref{fig_tire_Bootstrapping_parameter_Histogram}. As one can notice, such an algorithm belongs to non-Bayesian category. The accuracy of the model depends on the number of bootstrapped samples and the data-generating mechanism. The intricate peaks in the data are not captured which indicates that the model can further be refined. Bootstrap algorithm is simple and easy to implement, however Bayesian approaches give more control for the considered industrial application case and provide more mathematically robust formulations and diagnostic tools.
\section{Conclusions} \label{sec_conclusions}
In this research work, Bayesian metamodels are developed for vehicle broadband noises such as aerodynamic wind and tire-road interaction noise using measurement databases. For aerodynamic noise, two different Bayesian models are proposed which not just consider the available data but also rely on physical laws. The model parameters are interpretable where the domain-expert knowledge is exploited to define prior-distributions on them. The deterministic models are validated using $K$-fold cross validation with good accuracy and the Bayesian model assessment is done through various MCMC diagnostic tools. It is noted that, despite being computationally intensive, model with Gaussian basis functions produces better results, in terms of the distribution shape and Leave-One-Out cross validation statistic. However, for broadband masking noises overall estimators, detailed modelling is not usually desired and therefore BM1 can be used for fast computations. For tire-road noise, a non-Bayesian approach was tested and it was observed that Bayesian approaches provide more flexibility and control over the predicted responses.
\section{Acknowledgements}
This research work is funded by the European Commission’s H2020-Innovative training network (ITN) under the project ECODRIVE (Grant Nr. 858018) and we gratefully acknowledge the support of OpenLab Vibro-Acoustic-Tribology@Lyon, Laboratoire Vibrations Acoustique (LVA), INSA Lyon and the NVH department of Stellantis N.V.

\printbibliography

@article{durand2008structural,
	title={Structural-acoustic modeling of automotive vehicles in presence of uncertainties and experimental identification and validation},
	author={Durand, Jean-Fran{\c{c}}ois and Soize, Christian and Gagliardini, Laurent},
	journal={The Journal of the Acoustical Society of America},
	volume={124},
	number={3},
	pages={1513--1525},
	year={2008},
	publisher={Acoustical Society of America}
}

@article{kleijnen1975comment,
	title={A comment on Blanning's “Metamodel for sensitivity analysis: the regression metamodel in simulation”},
	author={Kleijnen, Jack PC},
	journal={Interfaces},
	volume={5},
	number={3},
	pages={21--23},
	year={1975},
	publisher={INFORMS}
}

@article{wang2007review,
	title={Review of metamodeling techniques in support of engineering design optimization},
	author={Wang, G Gary and Shan, Songqing},
	year={2007}
}

@article{kianifar2020performance,
	title={Performance evaluation of metamodelling methods for engineering problems: towards a practitioner guide},
	author={Kianifar, Mohammed Reza and Campean, Felician},
	journal={Structural and Multidisciplinary Optimization},
	volume={61},
	number={1},
	pages={159--186},
	year={2020},
	publisher={Springer}
}

@article{ibrahim_surrogate-based_2020,
	title = {Surrogate-Based Acoustic Noise Prediction of Electric Motors},
	volume = {56},
	issn = {1941-0069},
	pages = {1--4},
	number = {2},
	journaltitle = {{IEEE} Transactions on Magnetics},
	author = {Ibrahim, I. and Silva, R. and Mohammadi, M. H. and Ghorbanian, V. and Lowther, D. A.},
	date = {2020-02},
	note = {Conference Name: {IEEE} Transactions on Magnetics},
}

@article{kiani_comparative_2016,
	title = {A Comparative Study of Non-traditional Methods for Vehicle Crashworthiness and {NVH} Optimization},
	volume = {23},
	issn = {1134-3060, 1886-1784},
	pages = {723--734},
	number = {4},
	journaltitle = {Archives of Computational Methods in Engineering},
	shortjournal = {Arch Computat Methods Eng},
	author = {Kiani, Morteza and Yildiz, Ali R.},
	date = {2016-12},
}

@article{faes_recent_2020,
	title = {Recent Trends in the Modeling and Quantification of Non-probabilistic Uncertainty},
	volume = {27},
	issn = {1886-1784},
	pages = {633--671},
	number = {3},
	journaltitle = {Archives of Computational Methods in Engineering},
	shortjournal = {Arch Computat Methods Eng},
	author = {Faes, Matthias and Moens, David},
	date = {2020-07-01},
	langid = {english},
}

@article{zhang2012bayesian,
	title={Bayesian force reconstruction with an uncertain model},
	author={Zhang, Erliang and Antoni, Jerome and Feissel, Pierre},
	journal={Journal of Sound and Vibration},
	volume={331},
	number={4},
	pages={798--814},
	year={2012},
	publisher={Elsevier}
}

@article{pereira2015empirical,
	title={Empirical Bayesian regularization of the inverse acoustic problem},
	author={Pereira, Antonio and Antoni, Jerome and Leclere, Quentin},
	journal={Applied Acoustics},
	volume={97},
	pages={11--29},
	year={2015},
	publisher={Elsevier}
}

@article{cerrato2009automotive,
	title={Automotive sound quality--powertrain, road and wind noise},
	author={Cerrato, Gabriella},
	journal={Sound and Vibration},
	volume={43},
	number={4},
	pages={16--24},
	year={2009}
}

@article{oettle_automotive_2017,
	title = {Automotive aeroacoustics: An overview},
	volume = {231},
	shorttitle = {Automotive aeroacoustics},
	journaltitle = {Proceedings of the Institution of Mechanical Engineers, Part D: Journal of Automobile Engineering},
	shortjournal = {Proceedings of the Institution of Mechanical Engineers, Part D: Journal of Automobile Engineering},
	author = {Oettle, Nicholas and Sims-Williams, David},
	date = {2017-04-10},
}

@book{hastie2017generalized,
	title={Generalized additive models},
	author={Hastie, Trevor J and Tibshirani, Robert J},
	year={2017},
	publisher={Routledge}
}

@book{ruppert2003semiparametric,
	title={Semiparametric regression},
	author={Ruppert, David and Wand, Matt P and Carroll, Raymond J},
	number={12},
	year={2003},
	publisher={Cambridge university press}
}

@article{myers1989response,
	title={Response surface methodology: 1966--l988},
	author={Myers, Raymond H and Khuri, Andre I and Carter, Walter H},
	journal={Technometrics},
	volume={31},
	number={2},
	pages={137--157},
	year={1989},
	publisher={Taylor \& Francis}
}

@book{friedman2017elements,
	title={The elements of statistical learning: Data mining, inference, and prediction},
	author={Friedman, Jerome H},
	year={2017},
	publisher={springer open}
}

@article{breiman1992submodel,
	title={Submodel selection and evaluation in regression. The X-random case},
	author={Breiman, Leo and Spector, Philip},
	journal={International statistical review/revue internationale de Statistique},
	pages={291--319},
	year={1992},
	publisher={JSTOR}
}

@book{forrester2008engineering,
	title={Engineering design via surrogate modelling: a practical guide},
	author={Forrester, Alexander and Sobester, Andras and Keane, Andy},
	year={2008},
	publisher={John Wiley \& Sons}
}

@article{chen2008design,
	title={A design-driven validation approach using Bayesian prediction models},
	author={Chen, Wei and Xiong, Ying and Tsui, Kwok-Leung and Wang, Shuchun},
	year={2008}
}

@book{bolstad2009understanding,
	title={Understanding computational Bayesian statistics},
	author={Bolstad, William M},
	volume={644},
	year={2009},
	publisher={John Wiley \& Sons}
}

@book{gelman2013bayesian,
	title={Bayesian Data Analysis},
	author={Gelman, Andrew and Carlin, John B and Stern, Hal S and Dunson, David B and Vehtari, Aki and Rubin, Donald B},
	year={2013},
	publisher={CRC Press}
}

@book{neal1993probabilistic,
	title={Probabilistic inference using Markov chain Monte Carlo methods},
	author={Neal, Radford M},
	year={1993},
	publisher={Department of Computer Science, University of Toronto Toronto, ON, Canada}
}

@article{hoffman2014no,
	title={The No-U-Turn sampler: adaptively setting path lengths in Hamiltonian Monte Carlo.},
	author={Hoffman, Matthew D and Gelman, Andrew and others},
	journal={J. Mach. Learn. Res.},
	volume={15},
	number={1},
	pages={1593--1623},
	year={2014}
}

@book{murphy_machine_2012,
	location = {Cambridge, Massachusetts London, England},
	title = {Machine learning: a probabilistic perspective},
	isbn = {978-0-262-01802-9},
	series = {Adaptive computation and machine learning series},
	shorttitle = {Machine learning},
	pagetotal = {1071},
	publisher = {The {MIT} Press},
	author = {Murphy, Kevin P.},
	date = {2012},
	langid = {english},
}

@article{efron1979bootstrap,
	title={Bootstrap Methods: Another Look at the Jackknife},
	author={Efron, B},
	journal={The Annals of Statistics},
	volume={7},
	number={1},
	pages={1--26},
	year={1979},
	publisher={Institute of Mathematical Statistics}
}

@book{efron_introduction_1993,
	location = {Boston, {MA}},
	title = {An Introduction to the Bootstrap},
	isbn = {978-0-412-04231-7 978-1-4899-4541-9},
	publisher = {Springer {US}},
	author = {Efron, Bradley and Tibshirani, Robert J.},
	date = {1993},
	langid = {english}
}

@article{jordan_graphical_2004,
	title = {Graphical Models},
	volume = {19},
	issn = {0883-4237},
	number = {1},
	journaltitle = {Statistical Science},
	shortjournal = {Statist. Sci.},
	author = {Jordan, Michael I.},
	date = {2004-02-01},
	langid = {english},
}

@phdthesis{brogna2018probabilistic,
	title={Probabilistic Bayesian approaches to model the global vibro-acoustic performance of vehicles},
	author={Brogna, Gianluigi},
	year={2018},
	school={Lyon}
}

@book{martin2022bayesian,
	title={Bayesian Modeling and Computation in Python},
	author={Martin, Osvaldo A and Kumar, Ravin and Lao, Junpeng},
	year={2022},
	publisher={Chapman and Hall/CRC}
}

@article{salvatier2016probabilistic,
	title={Probabilistic programming in Python using PyMC3},
	author={Salvatier, John and Wiecki, Thomas V and Fonnesbeck, Christopher},
	journal={PeerJ Computer Science},
	volume={2},
	year={2016},
	publisher={PeerJ Inc.}
}

@article{kuvznar2012improving,
	title={Improving vehicle aeroacoustics using machine learning},
	author={Kuznar, Damjan and Mozina, Martin and Giordanino, Marina and Bratko, Ivan},
	journal={Engineering Applications of Artificial Intelligence},
	volume={25},
	number={5},
	pages={1053--1061},
	year={2012},
	publisher={Elsevier}
}

@article{vehtari2017practical,
	title={Practical Bayesian model evaluation using leave-one-out cross-validation and WAIC},
	author={Vehtari, Aki and Gelman, Andrew and Gabry, Jonah},
	journal={Statistics and computing},
	volume={27},
	number={5},
	pages={1413--1432},
	year={2017},
	publisher={Springer}
}

@article{vehtari2021rank,
	title={Rank-normalization, folding, and localization: An improved $\hat{R}$ for assessing convergence of MCMC},
	author={Vehtari, Aki and Gelman, Andrew and Simpson, Daniel and Carpenter, Bob and Burkner, Paul-Christian},
	journal={Bayesian analysis},
	volume={16},
	number={2},
	pages={667--718},
	year={2021},
	publisher={International Society for Bayesian Analysis}
}

@article{gabry2019visualization,
	title={Visualization in Bayesian workflow},
	author={Gabry, Jonah and Simpson, Daniel and Vehtari, Aki and Betancourt, Michael and Gelman, Andrew},
	journal={Journal of the Royal Statistical Society: Series A (Statistics in Society)},
	volume={182},
	number={2},
	pages={389--402},
	year={2019},
	publisher={Wiley Online Library}
}

@article{li2018literature,
	title={Literature review of tire-pavement interaction noise and reduction approaches},
	author={Li, Tan},
	journal={Journal of Vibroengineering},
	volume={20},
	number={6},
	pages={2424--2452},
	year={2018},
	publisher={JVE International Ltd.}
}

@article{mat_ali_wind_2018,
	title = {Wind noise from A-pillar and side view mirror of a realistic generic car model, {DriAver}},
	volume = {14},
	pages = {38},
	journaltitle = {International Journal of Vehicle Noise and Vibration},
	shortjournal = {International Journal of Vehicle Noise and Vibration},
	author = {Mat Ali, Mohamed Sukri and Jalasabri, Jafirdaus and Sood, Anwar and Mansor, Shuhaimi and Shaharuddin, Haziqah and Muhamad, Sallehuddin},
	date = {2018-01-01},
}

@book{soize2017uncertainty,
	title={Uncertainty quantification},
	author={Soize, Christian},
	year={2017},
	publisher={Springer}
}

@inproceedings{chen_multi-objective_2018,
	title = {Multi-Objective Optimization of Interior Noise of an Automotive Body Based on Different Surrogate Models and {NSGA}-{II}},
	eventtitle = {{WCX} World Congress Experience},
	pages = {2018--01--0146},
	author = {Chen, Shuming and Zhang, Jixiu},
	date = {2018-04-03},
}

@inproceedings{hoffman_robust_2003,
	title = {Robust Piston Design and Optimization Using Piston Secondary Motion Analysis},
	eventtitle = {{SAE} 2003 World Congress \& Exhibition},
	pages = {2003--01--0148},
	author = {Hoffman, Rebecca M. and Sudjianto, Agus and Du, Xiaoping and Stout, Joseph},
	date = {2003-03-03},
}

@inproceedings{zheng_design_2015,
	title = {The Design Optimization of Vehicle Interior Noise through Structural Modification and Constrained Layer Damping Treatment},
	eventtitle = {{SAE} 2015 World Congress \&  Exhibition},
	pages = {2015--01--0663},
	author = {Zheng, Ling and Fang, Zhanpeng and Tang, Zhongcai and Zhan, Zhenfei and Fu, Jiang-hua},
	date = {2015-04-14},
	langid = {english}
}

@inproceedings{park_efficient_2020,
	title = {Efficient Surrogate-Based {NVH} Optimization of a Full Vehicle Using {FRF} Based Substructuring},
	eventtitle = {{WCX} {SAE} World Congress Experience},
	pages = {2020--01--0629},
	author = {Park, Inseok and Papadimitriou, Dimitrios},
	date = {2020-04-14},
}

@inproceedings{barillon2012vibro,
	title={Vibro-acoustic variability of a body in white using Monte Carlo simulation in a development process},
	author={Barillon, F and Boubaker, MB and Mordillat, P and Lardeur, P},
	booktitle={Proceedings of International Conference on Noise and Vibration Engineering (ISMA), \& International Conference on Uncertainty in Structural Dynamics (USD), Leuven, Belgium},
	year={2012}
}
\end{document}